\documentclass[lettersize,journal]{IEEEtran}
\usepackage{amsmath,amsfonts}
\usepackage{algorithmic}
\usepackage{algorithm}
\usepackage{array}
\usepackage[caption=false,font=normalsize,labelfont=sf,textfont=sf]{subfig}
\usepackage{textcomp}
\usepackage{stfloats}
\usepackage{url}
\usepackage{verbatim}
\usepackage{graphicx}
\usepackage{cite}
\usepackage{bm}
\usepackage{amsmath}
\usepackage{multirow}
\usepackage{booktabs}
\usepackage{color}
\usepackage{soul} 
\usepackage{hyperref}
\usepackage{epstopdf}
\soulregister{\ref}7
\soulregister{\eqref}7
\soulregister{\cite}7
\newtheorem{Thm}{Theorem}

\newtheorem{Lem}{Lemma}

\newtheorem{Rem}{Remark}
\newtheorem{Ass}{Assumption}

\begin{document}

\title{Fixed-time Disturbance Observer-Based MPC Robust Trajectory Tracking Control
of Quadrotor}

\author{Liwen Xu, Bailing Tian, Cong Wang, Junjie Lu, Dandan Wang, Zhiyu Li, Qun Zong

\thanks{This work was supported in part by the National Natural Science Foundation of China under Grants 62273249, 62203415, 62373268, 62022060, 52304170, and 62373273. (Corresponding author: Zhiyu Li.)

Liwen Xu, Bailing Tian, Junjie Lu, Zhiyu Li and Qun Zong are with the Tianjin Key Laboratory of Intelligent Unmanned Swarm Technology and System, School of Electrical and Information Engineering, Tianjin University, Tianjin 300072, China (e-mail: xlw\_2000@tju.edu.cn; bailing\_tian@tju.edu.cn; lqzx1998@tju.edu.cn; lizhiyu@tju.edu.cn; zongqun@tju.edu.cn)

Cong Wang is with the College of Field Engineering, Army Engineering University of PLA, 210007, Nanjing, China (e-mail: lgd\_dolphin@139.com)

Dandan Wang is with Beijing Tianma Intelligent Control Technology Co., Ltd. Beijing 101399, China (e-mail: dandanwang0910@163.com)}}



\maketitle

\begin{abstract}
In this paper, a fixed-time disturbance observer-based model predictive control algorithm is proposed for trajectory tracking of quadrotor in the presence of disturbances. First, a novel multivariable fixed-time disturbance observer is proposed to estimate the lumped disturbances. The bi-limit homogeneity and Lyapunov techniques are employed to ensure the convergence of estimation error within a fixed convergence time, independent of the initial estimation error. Then, an observer-based model predictive control strategy is formulated to achieve robust trajectory tracking of quadrotor, attenuating the lumped disturbances and model uncertainties. Finally, simulations and real-world experiments are provided to illustrate the effectiveness of the proposed method.
\end{abstract}

\begin{IEEEkeywords}
Trajectory Tracking Control, Fixed-Time Disturbance Observer, Model Predictive Control, Quadrotor UAV.
\end{IEEEkeywords}

\section{Introduction}
\IEEEPARstart{Q}{uadrotor} Unmanned Aerial Vehicles (UAVs) have been utilized in diverse fields such as aerial delivery, inspection, rescue, and surveillance, owing to their remarkable capabilities, which include vertical take-off and landing, autonomous flight, and mapping \cite{10016632,9841590}. However, the control performance of quadrotor can be significantly affected by model uncertainty, aerodynamic drag force, and external disturbances, such as wind gusts and payloads with unknown mass. To ensure the safety and high performance of quadrotor in such scenarios, it is crucial to improve the trajectory tracking capability of quadrotor, especially in the presence of disturbances.

In recent years, numerous control methods have been proposed for trajectory tracking of quadrotor, which can be broadly categorized into linear control methods and nonlinear control methods. Linear control methods, such as classical PID \cite{bouabdallah2004pid,tayebi2006attitude} and linear quadratic regulator \cite{khatoon2014pid,ref4}, are relatively straightforward to implement and comprehend. Specifically, in \cite{tayebi2006attitude}, a PID-based controller that depends on the system model is proposed, providing exponential stability for the attitude system of quadrotor. However, due to the high nonlinear dynamics and strong coupling of quadrotor, nonlinear methods exhibit more potential to achieve outstanding performance, such as sliding mode control \cite{ref5_2,ref5_3,ref5_4}, backstepping control \cite{madani2006backstepping, guenard2008practical}, incremental nonlinear dynamic inversion (INDI) \cite{ref6,ref6_4}, active disturbance rejection control \cite{ref7} and neural networks-based control \cite{ref8}. On the other hand, the combination of different methods can significantly enhance the control performance of quadrotor. For instance, in \cite{ref5_4}, an adaptive method employed in the outer loop is integrated with a sliding mode control method employed in the inner loop, leading to substantial enhancement in robustness. In spite of the extensive development of robust techniques, the model predictive control (MPC) is widely acknowledged as a crucial tool for UAV control systems. The MPC method, by applying optimal control techniques within a specific context, adeptly addresses complex tasks through simultaneous consideration of multiple constraints and objectives via optimization.
\begin{figure}[tbp]
        \centering
        \includegraphics[width=3.4in]{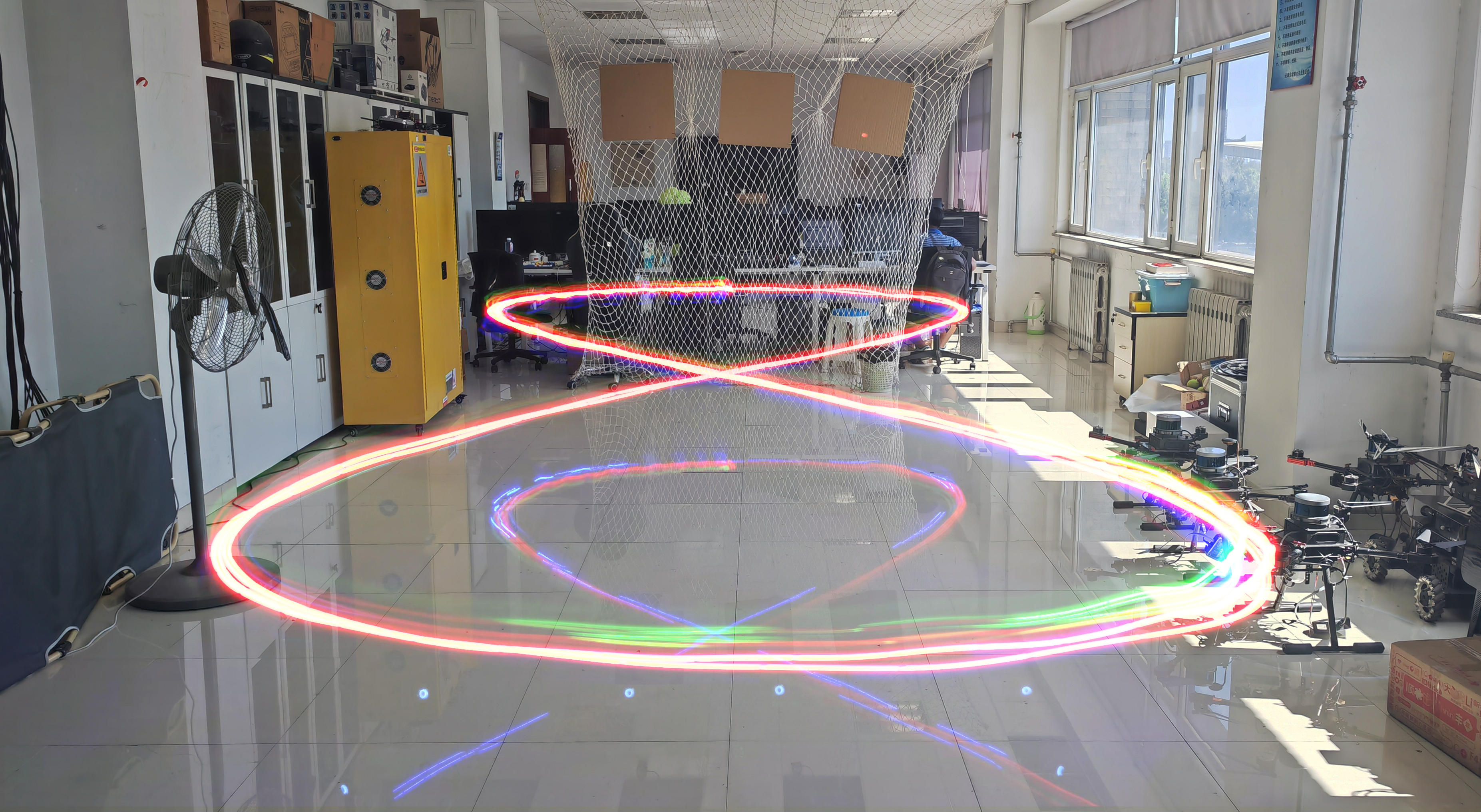}
        \caption{Trajectory visualization utilizing the proposed FxTDO-MPC algorithm in real-world experiments. More details can be found in the attached video at \href{https://youtu.be/O4BDXRdd0Do}{https://youtu.be/O4BDXRdd0Do} or \href{https://www.bilibili.com/video/BV11h4y1i7oF/}{https://www.bilibili.com/video/BV11h4y1i7oF/}.}
        \label{fig:trajectory_bling}
        \vspace{-0.5cm}
\end{figure}
There are existing literature on the application of MPC with disturbance estimation for quadrotor. In \cite{ref9,ref10}, a model predictive controller is proposed for position control of quadrotor. Additionally, augmented-state Kalman Filter disturbance estimators are designed to accurately estimate disturbances. In \cite{ref13}, a Gaussian Process method is utilized to model aerodynamic effect disturbances of quadrotor. However, offline learning procedure reduces the robustness and real-time performance of the controller. In \cite{ref24,ref25}, deep neural network-based methods are employed to predict disturbances and provide disturbance information for MPC. However, the effect relies on the training procedure and exhibits limited resilience to variations in the environment. In addition to the disturbance estimation methods mentioned above, some sliding mode disturbance observer techniques are integrated into the MPC framework. In \cite{zhang2023robust}, a finite-time sliding mode observer is employed within a robust model predictive control framework to estimate the additive disturbance for perturbed continuous-time systems. In \cite{oliveira2021disturbance}, a sliding mode disturbance observer is developed to estimate the lumped disturbances, which are then utilized by MPC for underwater-vehicle manipulators systems. Moreover, it has been shown that the convergence rate of disturbance observer plays an important role in constructing observer-based MPC control system \cite{yan2020non,mousavi2021integral}.

\textbf{Motivations:} In summary, the quadrotor is a strongly coupled and underactuated nonlinear system. It is challenging to achieve satisfactory performance using linear control methods for quadrotor \cite{khatoon2014pid,ref4}. In contrast, model predictive control (MPC), as a nonlinear optimization method, can effectively address the coupled multivariable nonlinear control problems by considering prediction model constraint. However, the success of MPC depends on the availability of a highly accurate prediction model \cite{ref10, ref13, ref24, ref25,yan2020non}. Therefore, the rapid provision of such an accurate model is an urgent problem to be solved \cite{ref13, ref24}. Motivated by these observations, a novel fixed-time disturbance observer (FxTDO) is proposed to accurately estimate the lumped disturbances with fast convergence. Subsequently, by integrating the estimation from the FxTDO with the nominal model, the FxTDO-based MPC (FxTDO-MPC) algorithm is developed to achieve robust trajectory tracking of quadrotor in the presence of disturbances. Fig.~\ref{fig:trajectory_bling} shows the trajectory of quadrotor utilizing the proposed FxTDO-MPC algorithm in real-world experiments. The main contributions of this paper can be summarized in the following two aspects.
 \begin{enumerate}[]
 	\item A novel multivariable FxTDO is proposed to estimate the lumped disturbances, and the bi-limit homogeneity and Lyapunov techniques are employed to ensure the convergence of estimation error within a fixed convergence time, independent of the initial estimation error.
 	\item By integrating the disturbance observations into the prediction model within the MPC framework, an FxTDO-MPC algorithm is developed to achieve robust trajectory tracking of quadrotor. Simulations and real-world experiments are presented to validate the effectiveness of the proposed algorithm. 
 \end{enumerate}

\begin{figure}[tbp]
        \centering
        \includegraphics[width=3.2in]{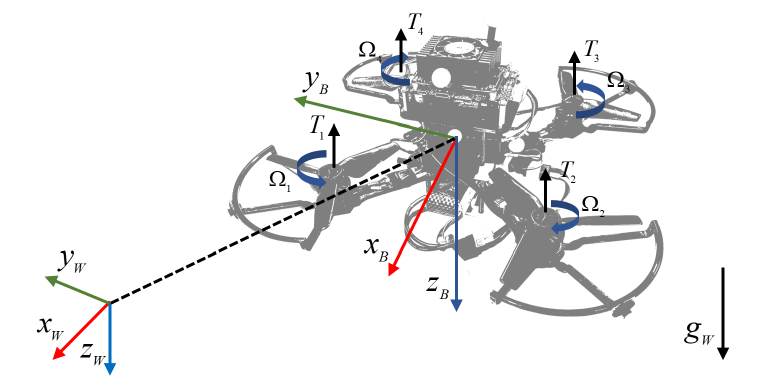}
        \caption{Quadrotor model with World frame $W$ and Body frame $B$. $T_i$ and $\Omega_i$ are respectively single thrust and speed of the individual rotors.} 
        \label{fig:Frame}
        \vspace{-0.5cm}
\end{figure}

\begin{figure*}[tbp]
        \centering
        \includegraphics[width=6.5in]{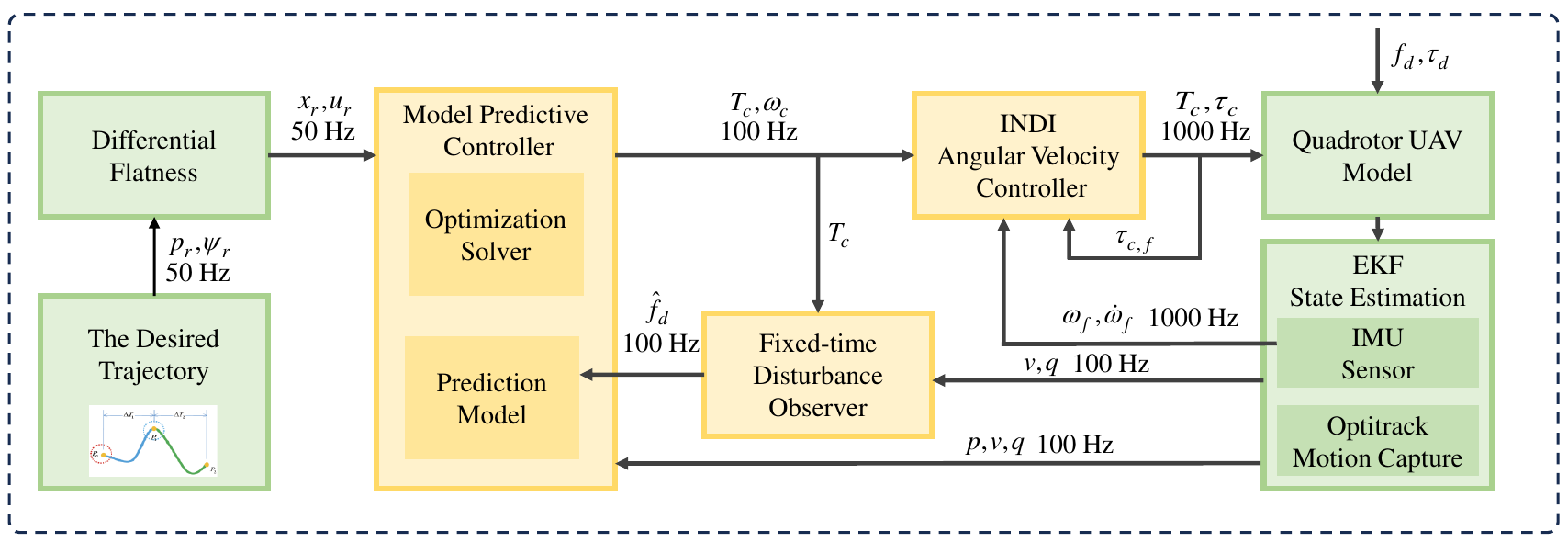}
        \caption{System overview of the proposed algorithm applied to quadrotor system. The frequencies of MPC and FxTDO are decided based on the computational time of FxTDO-MPC running on the onboard computer. Besides, the frequency of INDI angular velocity controller is selected based on the PX4 Autopilot User Guide. Additionally, the generating frequency of the desired trajectory is chosen to cooperate with the controller.}
        \label{fig:System}
        \vspace{-0.5cm}
\end{figure*}

The rest of this paper is organized as follows. In Section \uppercase\expandafter{\romannumeral2}, the dynamics of quadrotor and control objective are provided. The design of the FxTDO-MPC algorithm is introduced in Section \uppercase\expandafter{\romannumeral3}. In Section \uppercase\expandafter{\romannumeral4}, the results of simulations in Gazebo and real-world experiments are presented. Finally, Section \uppercase\expandafter{\romannumeral5} concludes this paper.

\section{Model and Control Objective Statement}
To describe the dynamics of quadrotor, two frames are defined: the World frame, denoted as $W$, and the Body frame, denoted as $B$, as illustrated in Fig.~\ref{fig:Frame}. The subscript of the variable i.e.$\left\{x_W,y_W,z_W\right\}$, indicates the specific frame where the variable is defined. $ x_W,y_W,z_W $ point north, east and down, and $  x_B,y_B,z_B $ point forward, right and down. Unless necessary, the subscript representing the coordinate frame is dropped throughout the paper for notational convenience. The frame $W$ is fixed on the ground. The frame $B$ is located at the center of mass of quadrotor. Four rotors, lying on the $xy$-plane of frame $B$, generate individual thrust ($T_1$-$T_4$) perpendicular to the $xy$-plane.

The model of quadrotor is defined as a rigid body with 6 degree-of-freedom (DoF). The dynamics of quadrotor can be described by:
\begin{subequations}\label{equation:Model}
        \begin{align}
            \dot{\bm{p}}&=\bm{v},\\
            \label{f_d}
            \dot{\bm{v}}&=-\frac{1}{m}\bm{R}_B^W(q)T_{c}\bm{e}_z^B+\bm{g}+\frac{1}{m}\bm{f}_d,\\
            \dot{\bm{q}}&=\frac{1}{2}\bm{q} \otimes
            \begin{bmatrix}
             0\\
            \bm{\omega}
            \end{bmatrix},\\
            \label{tau_d}
            \dot{\bm{\omega}}&=\bm{J}^{-1}[\bm{\tau}_c-\bm{\omega}\times\bm{J}\bm{\omega}]+\bm{J}^{-1}\bm{\tau}_d,
        \end{align}
\end{subequations}
where $\bm{p}=\left [  p_{x},p_{y},p_{z}\right ]^\mathsf{T} \in \mathbb{R} ^3$ and $\bm{v}=\left[v_{x},v_{y},v_{z}\right]^\mathsf{T}\in \mathbb{R} ^3$ are respectively position and velocity vectors of center of mass in the frame $W$. The symbol ${\mathsf{T}}$ represents the transpose. Unit quaternions, denoted as $\bm{q}=\left[q_{w},q_{x},q_{y},q_{z}\right]^\mathsf{T}\in \mathbb{R} ^4$ and satisfying $\left \| \bm{q} \right \|=1 $, represent the orintation of quadrotor in the frame $W$. $\bm{\omega}=\left[\omega_{x},\omega_{y},\omega_{z}\right]^\mathsf{T}\in \mathbb{R} ^3$ is angular velocity in the frame $B$, $m$ is the quadrotor mass, and $\bm{g}=\left[0,0,9.81 m/s^2\right]^{\mathsf{T}}$ denotes gravity vector in the frame $W$. $\bm{J} \in \mathbb{R}^{3\times3} $ is inertia matrix. The operator $\times$ represents the cross product. The vector $\bm{f}_d \in \mathbb{R}^3$ represents the lumped disturbance, which is caused by aerodynamic drag force and external disturbances, such as wind gusts and unknown payloads. The vector $\bm{\tau}_d \in \mathbb{R}^3$ represents unknown torque disturbance arising from inaccurate inertia matrix and external disturbances. $\bm{R}_B^W(q) \in \mathrm{SO}(3)$ denotes rotation matrices from frame $B$ to frame $W$:
\begin{equation}
          \bm{R}_B^W\!(q)\!=\!\begin{bmatrix}
          1\!-\!2q_y^2\!-\!2q_z^2\!& \!2(q_xq_y\!-\!q_wq_z) \!&\!2(q_wq_y\!+\!q_xq_z) \\
          2(q_xq_y\!+\!q_wq_z)\!&\! 1\!-\!2q_x^2\!-\!2q_z^2 \!&\!2(q_yq_z\!-\!q_wq_x) \\
          2(q_xq_z\!-\!q_wq_y)\!& \!2(q_wq_x\!-\!q_yq_z) \!&\!1\!-\!2q_x^2\!-\!2q_y^2
        \end{bmatrix}\!.\!
\end{equation}

Additionally, $T_c \in \mathbb{R}$ and $\bm{\tau}_c=\left[\tau_x,\tau_y,\tau_z\right]^\mathsf{T} \in \mathbb{R}^3$, as (\ref{equation:T and tau}) shows, are respectively the collective thrust command and the body torque command generated from four rotors. $\bm{e}_z^B=\left[0,0,1\right]^\mathsf{T}$ means z-direction of the frame $B$. Consequently, $\bm{R}_B^W(q)T_{c}\bm{e}_z^B$ means rotating $\left[0,0,T_c\right]$ from the frame $B$ to the frame $W$. Operator $\otimes$ is defined as:
\begin{equation}
        \bm{q} \otimes
        \begin{bmatrix}
         0\\
        \bm{\omega}
        \end{bmatrix}=
        \begin{bmatrix}
      0& -w_x & -w_y &-w_z\\
      w_x& 0 & w_z &-w_y \\
      w_y& -w_z &  0&w_x \\
      w_z& w_y &-w_x  &0
    \end{bmatrix}\bm{q}.
\end{equation}

Finally, $T_c$ and $\bm{\tau}_c$ can be computed based on the single thrust of each rotor:
\begin{equation}
        \label{equation:T and tau}
        T_c = \sum T_i   \quad, \quad \bm{\tau}_c=
        \begin{bmatrix}
     d_y(-T_0-T_1+T_2+T_3)\\d_x(-T_0+T_1+T_2-T_3)
     \\c_\tau(-T_0+T_1-T_2+T_3)
    \end{bmatrix},
\end{equation}
where $d_x$, $d_y$ are the rotor displacements and $c_\tau$ is the rotor drag torque constant.

The main objective of the presented work is to develop a fixed-time disturbance observer-based MPC methodology that computes appropriate $T_c$ and $\bm{\tau}_c$ to accurately track the desired trajectory while effectively compensating for disturbances. 
To achieve the control objectives, the following assumption is required.
\begin{Ass}
        \label{Ass:f_d}
	Suppose that the lumped disturbance vector $\bm{f}_d$ in \eqref{f_d} is continuous differentiable, and there exist some positive constants $\delta$ and $\overline{\delta}$ such that $\left\| \bm{f}_d \right\| \le \delta$ and $\left\| \dot{\bm{f}_d} \right\| \le \overline{\delta}$.
\end{Ass}

\section{Design of FxTDO-MPC Algorithm and INDI Angular Velocity Controller}
In this section, an FxTDO-MPC algorithm is proposed. First, an FxTDO is designed to estimate the lumped disturbances. Then, a model predictive controller is developed to track the desired trajectory while compensating for the lumped disturbances. Finally, an INDI angular velocity controller is adopted to deal with unknown torque disturbance. The architecture of the integrated controller and observer is provided in Fig.~\ref{fig:System}.

\subsection{Related Lemmas} 
\begin{Lem} \cite{cruz2021high}
        Let $\eta: {\mathbb{R}}^n \to {\mathbb{R}}$ and $\gamma : {\mathbb{R}}^n \to {\mathbb{R}}_+ $ be two upper semicontinuous, single-valued bl-homogeneous functions, with the same weights, degrees and approximating functions. Suppose that $ \forall x \in {\mathbb{R}}^n$, $\left\lbrace  x \in {\mathbb{R}}^n \backslash \{0\}: \gamma (x) =0 \right\rbrace \subseteq \left\lbrace  x \in {\mathbb{R}}^n \backslash \{0\}: \eta (x) < 0 \right\rbrace $. Then, there exists a real number $\lambda^*$ such that, for all $\lambda \ge \lambda^*$, $\eta (x) - \lambda \gamma (x) < 0 $.
\end{Lem}

\begin{Lem} \cite{cruz2021high}
        Consider a homogeneous in the bi-limit set-valued vector field $\bm{f}: {\mathbb{R}}^n \to {\mathbb{R}}^n$, which is upper semicontinuous, with associated triples $(\bm{r}_0, k_0, \bm{f}_0)$ and $(\bm{r}_\infty, k_\infty, \bm{f}_\infty)$ such that the origins of the system \eqref{Eq1} are globally asymptotically stable equilibria. Let $m_0$ and $m_\infty$ be real numbers such that $ m_0 > \max \{r^0_i \}$ and $ m_\infty > \max \{r^\infty_i \}$. Then there exists a $C^1$, positive definite function $V(\bm{x})$ such that the derivative of  $V(\bm{x})$ is bl-homogeneous and negative definite. Moreover, if $k_0 \le k_\infty$ there exist positive constants $\kappa_0 > 0$ and $\kappa_\infty> 0$ such that the inequality hold for all $\bm{x} \in {\mathbb{R}}^n$, $\dot{V} (\bm{x}) \le - \kappa_0  V^\frac{m_0 + k_0}{m_0}(\bm{x}) - \kappa_\infty  V^\frac{m_\infty + k_\infty}{m_\infty}(\bm{x}) $.
\end{Lem}

\subsection{Fixed-Time Disturbance Observer Design}
Firstly, an FxTDO is introduced to estimate the lumped disturbance $\bm{f}_d$ in (5), guaranteeing the convergence of the estimation error within a fixed time T, i.e., the estimation error converges globally in finite time, and the convergence time is globally bounded by a positive constant T, independent of the initial estimation error \cite{andrieu2008homogeneous}. Consider the quadrotor dynamics (\ref{f_d}), which can be rewritten as the following multivariable first-order system:
\begin{equation}\label{Eq1}
        \dot {\bm{z}}_1 = \bm{T} + \bm{f}_d,
\end{equation}
where $\bm{z_1} = m \bm{v}$, $\bm{T} = m \bm{g} - \bm{R}_B^W(q)T_{c}\bm{e}_z^B$ is the virtual control input. Inspired by our previous work in \cite{tian2017fixed}, by including a discontinuous term $\lceil\bm{e_1}\rfloor ^{0}$,  the following improved observer is designed to estimate the lumped disturbance $\bm{f}_d$. 
\begin{equation}
	\begin{aligned}\label{Eq2}
	 \bm{\dot{\hat{z}}_1} &= \bm{\hat{f}_d} +  \bm{T} + L_1 \bm{\phi_1}(\bm{e_1}),\\
	\bm{\dot{\hat{f}}_d} &=  L_2 \bm{\phi_2}(\bm{e_1}),
	\end{aligned}
\end{equation}
with $\bm{e_1} = \ \bm{z_1} - \bm{\hat{z}_1}$, and $\hat{f}_d$ is the estimation of $\bm{f}_d$. $L_1$ and $L_2$ are the gains to be designed to ensure the stability of the observer, and the functions $\bm{\phi_1}$ and $\bm{\phi_2}$ are given by:
\begin{equation}\label{Eq3}
	\begin{aligned}
		\bm{\phi_1}(\bm{e_1}) &= k_1 \lceil\bm{e_1}\rfloor^{\frac{1}{2}} + {k'_1} \lceil\bm{e_1}\rfloor + k''_1 \lceil\bm{e_1}\rfloor^{\frac{1}{1 - d_{\infty}}},\\
		\bm{\phi_2}(\bm{e_1}) &= k_2 \lceil\bm{e_1}\rfloor^{0} + {k'_2} \lceil\bm{e_1}\rfloor + k''_2 \lceil\bm{e_1}\rfloor^{\frac{1 + d_{\infty} }{1 - d_{\infty}}},
	\end{aligned}
\end{equation}
where $d_{\infty} \in \left(  0, 1 \right) , k_i> 0, {k'_i}>0, {k''_i}>0, i=(1,2)$. Note that $\bm{\phi_1}(\bm{e_1})$ and $\bm{\phi_2}(\bm{e_1})$ are bi-homogeneous, with the degree of homogeneity at the 0-limit being $-1$ and at the $\infty$-limit being $d_\infty$. The comprehensive definition of bi-homogeneity can refer to \cite{andrieu2008homogeneous} and \cite{moreno2021arbitrary}. 

Subtracting \eqref{Eq2} from \eqref{Eq1} yields the following error dynamics
\begin{equation}\label{Eq4}
	\begin{aligned}
		\bm{\dot e_1} &=  - L_1 \left( \bm{\phi_1}(\bm{e_1}) - \bm{e_2}\right) ,\\
		\bm{\dot e_2} &= -\frac{L_2}{L_1} \left( \bm{\phi_2}(\bm{e_1}) - \frac{\dot{\bm{f}_d}}{L_2}\right),
	\end{aligned}
\end{equation}
with $\bm{e_2} = \left( \bm{f}_d - \bm{\hat{f}_d}\right) /L_1$. 

The main result of the FxTDO is summarized in the following theorem.
\begin{Thm}
For system \eqref{Eq1} with Assumption 1, select $0< d_{\infty} < 1$ and choose arbitrary positive gains  $k_i> 0, {k'_i}>0, {k''_i}>0$ for  $i=1,2$. Then, there exist appropriate gains $L_1$ as specified in  \eqref{L_1} and $L_2$ satisfying
\begin{equation}\label{L_2}
      L_2 > \overline{\delta}/k_2,
\end{equation}
such that the observer error dynamics given in \eqref{Eq4} are fixed-time stable, i.e., the observer \eqref{Eq2} is able to estimate the lumped disturbance  $\bm{f}_d$ in a fixed time $\overline{T}$, i.e., $\bm{\hat{f}_d} = \bm{f}_d$ for $t>\overline{T}$.
\end{Thm}

\textbf{Proof:}	Consider the bi-homogeneous Lyapunov function derived in \cite{moreno2021arbitrary} 
\begin{equation}\label{Eq5}
	V(\bm{e}) = V_1(\bm{e_1}, \bm{e_2}) + V_2(\bm{e_2}),
\end{equation}	
where the functions $V_1$ and $V_2$ are defined as
\begin{equation}\label{Eq6}
	\begin{aligned}
	&V_1(\bm{e_1}, \bm{e_2}) \!=\! \beta_0 \left( \frac{2}{p_0} \left\| \bm{e_1} \right\|^{\frac{p_0}{2}} \!-\! \bm{e_1}^\mathsf{T} \lceil\bm{\xi}\rfloor^{\frac{p_0 - 2}{2}} \!+\! \frac{p_0\!-\!2}{2} \left\| \bm{\xi} \right\|^{\frac{p_0}{2}} \right) \\
	 &\!+\! \beta_\infty \!\left( \frac{1 \!-\! d_\infty}{p_\infty}\! \left\| \bm{e_1}\! \right\|^{\frac{p_\infty\!}{2}} \!-\! \bm{e_1}^\mathsf{T} \!\lceil\bm{\xi}\rfloor^\frac{p_\infty \!-1\! +\! d_\infty\!}{1-d_\infty} \! +\!  \frac{\!p_\infty\! \!-\! 1 \!+\! d_\infty\!}{p_\infty} \left\| \bm{\xi} \right\|^{\frac{p_\infty}{\!1\!-\!d_\infty\!}\!}\!\right)
	\end{aligned}
\end{equation}	
\begin{equation}\label{Eq7}
	\begin{aligned}
		V_2(\bm{e_2}) =  \frac{\beta'_0}{p_0} \left\| \bm{e_2} \right\|^{p_0} + \frac{\beta'_\infty}{p_\infty} \left\| \bm{e_2} \right\|^{p_\infty},
	\end{aligned}
\end{equation}	
with  $0 < p_\infty \le \max \left\lbrace 1, 3(1 - d_\infty)/2 \right\rbrace, 1 < p_0 \le 2 p_\infty / (1-d_\infty)$, and $\beta_0, \beta_\infty, \beta'_0, \beta'_\infty$ being arbitrary positive real numbers.  $\bm{\xi}$ is defined as $\bm{\xi} = \bm{\phi_1}^{-1}(\bm{e_2})$.  According to Young's inequality in \cite{hardy1952inequalities}, it can be concluded that $V_1(\bm{e_1}, \bm{e_2}) \ge 0 $, which ensures the positive definiteness of Lyapunov function $V(\bm{e})$ in \eqref{Eq5}. 

The derivative of $V(\bm{e})$ along system \eqref{Eq4} can be written as
\begin{equation}\label{Eq8}
	\begin{aligned}
		\dot V(\bm{e}) = & W(\bm{e}),\\
		W(\bm{e}) \!=&\!  - \! \bm{g_1} \!L_1 \! \left( \bm{\phi_1}(\!\bm{e_1}\!) \!-\! \bm{e_2}\right)  \!-\!  \frac{\left( \bm{g_2} \!+\! \bm{\sigma_1\!}\right)L_2}{L_1}  \left( \bm{\phi_2}(\bm{e_1}\!) \!-\! \frac{\dot{\bm{f}_d}}{L_2}\right),
	\end{aligned}
\end{equation}	
where 
\begin{equation}\label{Eq9}
		\begin{aligned}
 \bm{g_1} &= \beta_0 \left( \bm{e_1}^\mathsf{T} \left\| \bm{e_1} \right\|^{\frac{p_0 - 4}{2}} - \bm{\xi}^\mathsf{T} \left\| \bm{\xi} \right\|^{\frac{p_0 - 4}{2}} \right)\\
&+ \beta_\infty \left( \bm{e_1}^\mathsf{T} \left\| \bm{e_1} \right\|^{\frac{p_\infty}{1 - d_\infty} - 2 } - \bm{\xi}^\mathsf{T} \left\| \bm{\xi} \right\|^{\frac{p_\infty}{1 - d_\infty} - 2 } \right),
	\end{aligned}
\end{equation}	
\begin{equation}\label{Eq10}
	\begin{aligned}
		\bm{\sigma_1} &\!=\! \beta_0  \left\| \bm{\xi}\right\|^{\frac{p_0 - 4}{2}} \frac{\partial \bm{\xi} }{\partial \bm{e_2}} \left( \bm{\xi}^\mathsf{T} \!-\! \bm{e_1}^\mathsf{T} \!+\! \frac{p_0 \!-\! 4}{2} (\bm{\xi}^\mathsf{T}  \!-\! \frac{\bm{e_1}^\mathsf{T} \bm{\xi} \bm{\xi}^\mathsf{T} }{\left\| \bm{\xi}\right\|^2})\right)  \\
		&+\! \beta_\infty  \left\| \bm{\xi}\right\|^{\frac{p_\infty}{1 - d_\infty} - 2} \frac{\partial \bm{\xi} }{\partial \bm{e_2}} \left( \bm{\xi}^\mathsf{T} \!-\! \bm{e_1}^\mathsf{T} \right) \\
                &+\! \beta_\infty  \left\| \bm{\xi}\right\|^{\frac{p_\infty}{1 - d_\infty} - 2} \frac{\partial \bm{\xi} }{\partial \bm{e_2}} \left( \frac{p_\infty \!-\! 2 \!+\! 2 d_\infty}{1 \!-\! d_\infty} (\bm{\xi}^\mathsf{T}  \!-\! \frac{\bm{e_1}^\mathsf{T} \bm{\xi} \bm{\xi}^\mathsf{T} }{\left\| \bm{\xi}\right\|^2})\right),
	\end{aligned}
\end{equation}	
\begin{equation}\label{Eq11}
	\begin{aligned}
		\bm{g_2} = \beta'_0 \bm{e_2}^\mathsf{T} \left\| \bm{e_2} \right\|^{p_0 - 2} + \beta'_\infty \bm{e_2}^\mathsf{T} \left\| \bm{e_2} \right\|^{p_\infty - 2}.
	\end{aligned}
\end{equation}	

Now, we are in the position to demonstrate the negative definiteness of $W(\bm{e})$. To this end, consider the structure of $W(\bm{e})$, which is restricted to the hypersurfaces
\begin{equation}\label{eq18}
	\begin{aligned}
		\bm{Z_1} = \left\lbrace \bm{\phi_1}(\bm{e_1}) = \bm{e_2} \right\rbrace.
	\end{aligned}
\end{equation}	

\begin{figure}[tbp]
        \centering
        \includegraphics[width=3.0in]{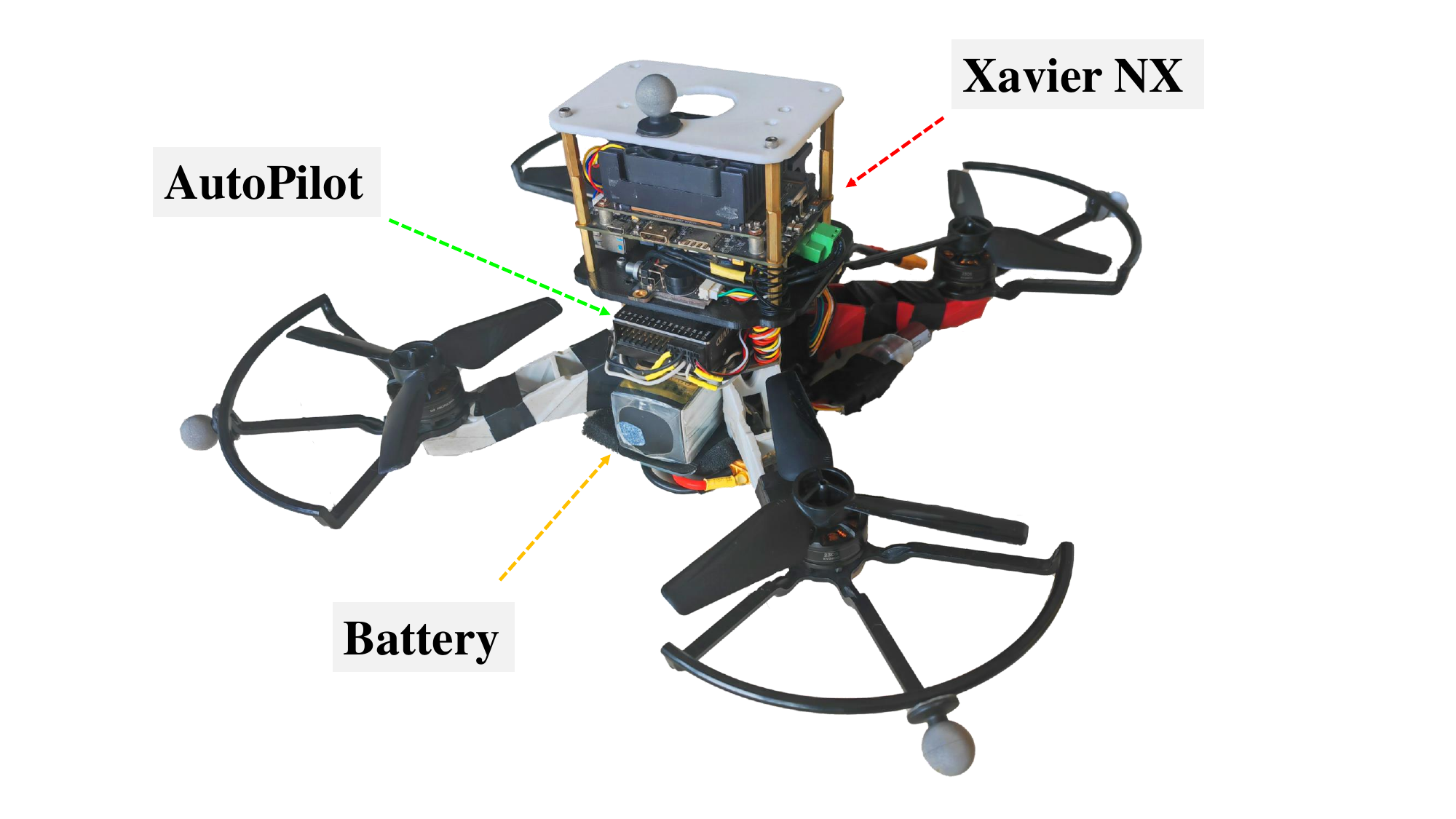}
        \caption{The quadrotor used for real-world experiments.}
        \label{fig:real quadrotor}
        \vspace{-0.5cm}
\end{figure}

Note that on the hypersurface $\bm{Z_1}$, we have $\bm{e_1} =  \bm{\phi_1}^{-1}(\bm{e_{2}})= \bm{\xi}$. Therefor, we can conclude form \eqref{Eq9} and \eqref{Eq10} that  $\bm{g_1}, \bm{\sigma_1} \in \{0\}$  for $\bm{Z_1}$. Then, let $W_{\bm{Z_1}}$ denotes the value of $W(e)$ restricted to the manifold $\bm{Z_1}$, i.e., by substituting $\bm{g_1}=\bm{\sigma_1}=0$ into \eqref{Eq8}, $W_{\bm{Z_1}}$ can be written as
\begin{equation}\label{eq19}
	\begin{aligned}
		W_{\bm{Z_1}} &=   - \frac{\bm{g_2} L_2}{L_1} \left(  \bm{\phi_2}(\bm{e_1}) - \frac{\dot{\bm{f}_d}}{L_2} \right)\\
		& = -\left( \beta'_0  \left\| \bm{e_2} \right\|^{p_0 - 2} + \beta'_\infty  \left\| \bm{e_2} \right\|^{p_\infty - 2}\right)*\\
		& \frac{L_2}{L_1} \bm{e_2}^\mathsf{T} \left(k_2 \lceil\bm{e_1}\rfloor^{0} + {k'_2} \lceil\bm{e_1}\rfloor + k''_2 \lceil\bm{e_1}\rfloor^{\frac{1 + d_{\infty} }{1 - d_{\infty}}} - \frac{\dot{\bm{f}_d}}{L_2} \right).
	\end{aligned}
\end{equation}	

Taking into account the properties of the multivariable signum function $\bm{x}^\mathsf{T} \mathrm{sign}(\bm x) = \left\|  \bm x \right\|$, and considering that $\bm{e_2}= k_1 \lceil\bm{e_1}\rfloor^{\frac{1}{2}} + {k'_1} \lceil\bm{e_1}\rfloor + k''_1 \lceil\bm{e_1}\rfloor^{\frac{1}{1 - d_{\infty}}}$ on the hypersurface $\bm{Z_1}$, with the definition of $W^*_{\bm{Z_1}} = \max \left\lbrace W_{\bm{Z_1}} \right\rbrace $, $W^*_{\bm{Z_1}}$ can be expressed as
\begin{equation}\label{eq20}
	\begin{aligned}
		W^*_{\bm{Z_1}} \!=\! -\!\left( \!\beta'_0 \! \left\| \bm{e_2} \right\|^{p_0 - 2}\! + \!\beta'_\infty \!\left\| \bm{e_2} \right\|^{p_\infty - 2}\right)  \frac{L_2}{L_1}  \left\|\! \bm{e_2} \right\| \!\left(k_2  \!-\! \frac{\overline{\delta}}{L_2}\! \right).
	\end{aligned}
\end{equation}	

Therefor, given that  $L_2$ satisfies the inequity $L_2 > \overline{\delta}/k_2$ in \eqref{L_2}, function $W^*_{\bm{Z_1}}$ is negative definite. Moreover, from Lemma 1, we can conclude that there exists a positive value of $L_1$ satisfying 
\begin{equation}\label{L_1}
		\begin{aligned}
	L_1 > &\mathop {\max }\limits_{\left( {\bm{e_1},\bm{e_2} \in \mathbb{R} ^3} \right)} \left\lbrace  \bm{\chi} \left( \bm{e_1},\bm{e_2}\right)  \right\rbrace ,\\
	\bm{\chi} \left( \bm{e_1},\bm{e_2}\right)  = &\left(  \frac{\left(  \bm{g_2} \!+\! \bm{\sigma_1}\right)  \left(  \overline{\delta} - L_2 \bm{\phi_2}(\bm{e_1}) \right)}{ \bm{g_1} \left( \bm{\phi_1}(\bm{e_1}) \!-\! \bm{e_2}\right)} \right)^{\frac{1}{2}},
		\end{aligned}
\end{equation}
such that $W(\bm{e})$ is negative definite. Note that $\bm{\chi} \left( \bm{e_1},\bm{e_2}\right)$ is an upper semicontinuous, homogeneous function, and it has a maximum achieved on the homogeneous sphere \cite{cruz2016lyapunov}. Then, according to Lemma 2, the Lyapunov function $V(\bm{e})$ satisfies \eqref{eq21} for some positive constants $\eta_0, \eta_\infty$. 
\begin{equation}\label{eq21}
		\dot V(\bm{e}) \le - \eta_0 V^{\frac{p_0 -1}{p_0}}(\bm{e}) - \eta_\infty V^{\frac{p_\infty + d_\infty}{p_\infty}}(\bm{e}).
\end{equation}	

Finally, we can obtain the estimation of the fixed convergence time $\overline{T}$
\begin{equation}\label{eq22}
		\overline{T} \le \frac{p_0}{\eta_\infty} \left( \frac{p_\infty}{p_0 d_\infty} + 1\right) \left( \frac{\eta_0}{\eta_\infty} \right)^ {- \frac{1}{\frac{p_\infty}{p_0 d_\infty} + 1} },
\end{equation}	
which is independent of the initial estimation error \cite{polyakov2011nonlinear}. As a result, the fixed-time stable of the observer error dynamics \eqref{Eq4} is guaranteed. This completes the proof.
\subsection{Trajectory Tracking Controller Design}
To achieve the objective of accurate trajectory tracking in the presence of disturbances, a cascade control strategy is proposed, which is composed of two components: an FxTDO-based MPC controller and an INDI angular velocity controller. The FxTDO-based MPC controller is proposed to track the desired trajectory while simultaneously compensating for the lumped disturbance $\bm{f}_d$. In addition, the INDI angular velocity controller is developed to track the commands generated by the FxTDO-based MPC controller and deal with unknown torque disturbance $\bm{\tau}_d$.

\subsubsection{FxTDO-based MPC controller}
By integrating the estimation of FxTDO into the prediction model, an FxTDO-based MPC controller is developed to achieve robust trajectory tracking of quadrotor. The prediction model employed within the MPC problem is formulated by

\begin{subequations}\label{equation:h_Model}
        \begin{align}
            \dot{\bm{p}}&=\bm{v},\\
            \dot{\bm{v}}&=-\frac{1}{m}\bm{R}_B^W(q)T_{c}\bm{e}_z^B+\bm{g}+\frac{1}{m}\hat{\bm{f}}_d,\\
            \dot{\bm{q}}&=\frac{1}{2}\bm{q} \otimes
            \begin{bmatrix}
             0\\
            \bm{\omega}_c
            \end{bmatrix},
        \end{align}
\end{subequations}
where $\hat{\bm{f}}_d$ is the estimated value of the lumped disturbance provided by FxTDO. During each iteration of solving the MPC problem, the estimated disturbance $\hat{\bm{f}}_d$ is assumed to be constant along the prediction horizon. 

The controller generates control commands by solving an optimal problem in a receding horizon fashion. To formulate the optimization problem of FxTDO-based MPC controller, the state vector is defined as:
\begin{equation}\label{equation:x}
        \bm{x}=\left[\bm{p},\bm{v},\bm{q}\right]^\mathsf{T},
\end{equation}
and control command is defined:
\begin{equation}\label{equation:u}
    \bm{u}=\left[T_c,\bm{\omega}_c\right]^\mathsf{T}=\left[T_c,\omega_{x,c},\omega_{y,c},\omega_{z,c}\right]^\mathsf{T},
\end{equation}
where $\bm{\omega}_c = \left[\omega_{x,c},\omega_{y,c},\omega_{z,c}\right]^\mathsf{T}$ is the angular velocity command sent to the inner-loop controller. 

To formulate the cost function of the optimization problem, which is defined as the cost of states error and control commands error, the reference states and control commands are computed using the concept of differential flatness for quadrotor \cite{ref14}. For trajectory tracking of quadrotor, position $\bm{p}$ and yaw angle $\psi$ are selected to be flat outputs, which means that the desired trajectory $\bm{\varepsilon}_r$ is defined as $\bm{\varepsilon}_r=\left[\bm{p}_r,\psi_r\right]^\mathsf{T}$. The system is discretized into $N$ steps over time horizon $T$ of size $dt = T/N$, yielding a constrained nonlinear optimization problem:
\begin{equation}
        \begin{array}{cl}\label{equation:ocp}
        \min\limits_{\bm{u}} & \sum_{k=0}^{N-1}\!\left(\Delta \bm{x}_k^\mathsf{T}\bm{Q}\Delta \bm{x}_k\!+\!\Delta \bm{u}_k^\mathsf{T}\bm{R}\Delta \bm{u}_k\right)\!+\!\Delta \bm{x}_N^\mathsf{T}\bm{P}\Delta \bm{x}_N\\
        \text { s.t. } & \dot{\bm{x}}=\bm{h}\left(\bm{x}, \bm{u} \right), \bm{x}_{0}=\bm{x}_{init},\bm{u}\in\mathbb{U},\bm{x} \in \mathbb{X},\bm{x}_{N} \in \mathbb{X}_N
        \end{array}
\end{equation}
where $\Delta \bm{x}_k = \bm{x}_k-\bm{x}_{k,r}$ and $\Delta \bm{u}_k = \bm{u}_k-\bm{u}_{k,r}$ are respectively the errors of states and control commands, $\bm{x}_{k,r}$ and $\bm{u}_{k,r} $ are respectively the reference state vector and the reference control command vector computed from the desired trajectory. $\boldsymbol{Q}\succeq   0$ and $\boldsymbol{R}\succeq   0$ are positive definite weight matrixes respectively of states error and control commands error, while $\boldsymbol{P}\succeq   0$ is positive definite weight matrix of terminal states. $\bm{h}\left(\bm{x},  \bm{u} \right)$ is the prediction model represented in \eqref{equation:h_Model}. $\mathbb{U}$, $\mathbb{X}$ and $\mathbb{X}_N$ are constraints of respectively control commands, states and terminal states. The quadratic optimization problem (\ref{equation:ocp}) is constructed as a multiple shooting scheme, and a real-time iterative scheme based on the sequential quadratic programming algorithm is employed to solve it. Furthermore, the ACADO \cite{ref15} toolkit and qpOASES \cite{ref16} solver are utilized to solve this optimization problem.

\begin{figure}[t]
        \centering
        \includegraphics[width=2.0in]{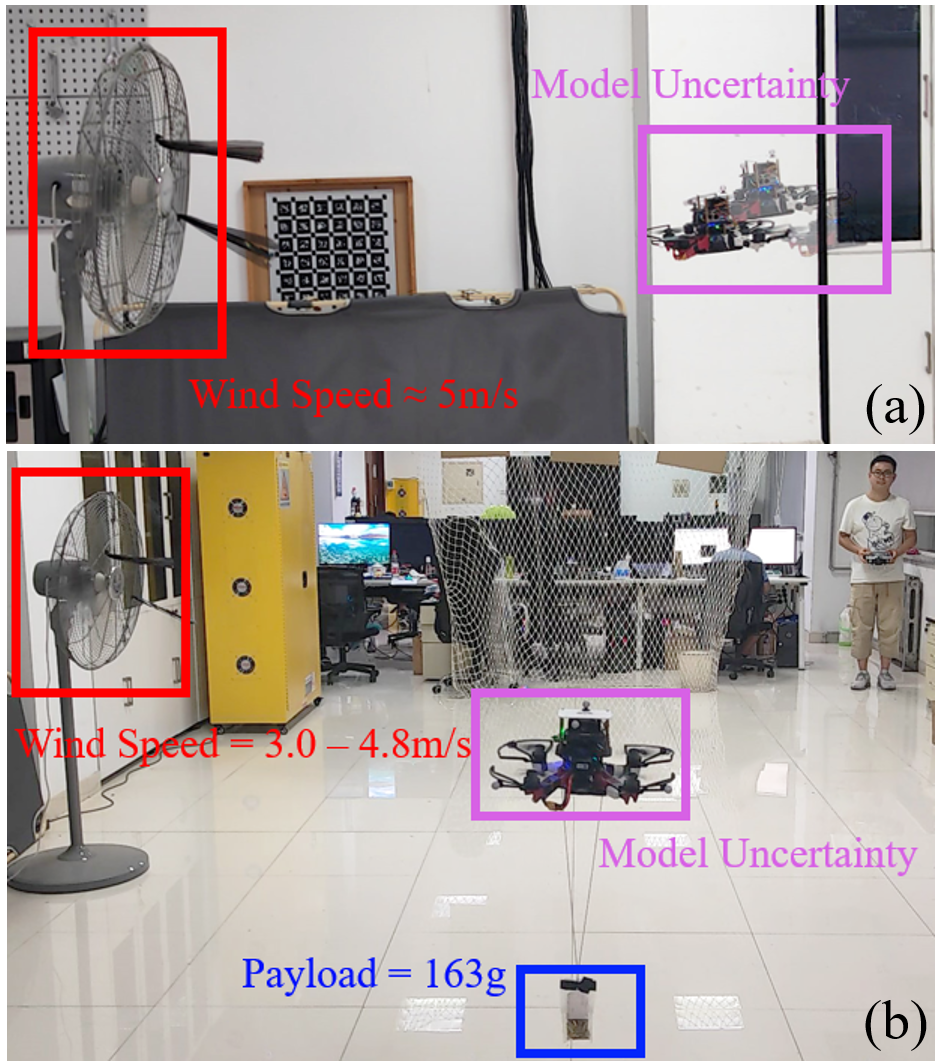}
        \caption{Disturbances in real-world experiments. (a) Disturbances in hovering tests \textit{(Scenario 1)}. (b) Disturbances in trajectory tracking tests \textit{(Scenario 2)}.}
        \label{fig:disturbance_setup}
        \vspace{-0.5cm}
\end{figure}

In what follows, the proof of the recursive feasibility of the proposed MPC and the stability of the combined system are provided based on the work in \cite{mayne2000constrained,fang2022model}. Firstly, for brevity, the optimization problem \eqref{equation:ocp} can be rewritten as the problem of steering the state to the origin in the following:


\begin{equation}\label{equation:ocp_proof}
        \begin{array}{cl}
        &\begin{array}{cl}
        V_N^0(x(k)) =\min\limits_{\bm{u}} &\sum_{i=0}^{N-1}l(x(k+i \mid k), u(k+i \mid k))\\
         &  + V_f(x(k+N \mid k)),\\
        \end{array}\\
         &\text { s.t. }
         \begin{array}{cl}
                &x(k+i+1 \mid k) = f(x(k+i \mid k),u(k+i \mid k)),\\
                & x(k+i \mid k)\in \mathbb{X}, u(k+i \mid k)\in \mathbb{U},\\
                & x(k+N \mid k)\in \mathbb{X}_N, x(k \mid k) = x(k),
        \end{array}\\
        
        \end{array}
\end{equation}
where $f(x(k+i \mid k),u(k+i \mid k)) = x(k+i \mid k) + h(x(k+i \mid k),u(k+i \mid k))*dt$. Solving the problem \eqref{equation:ocp_proof} yields the optimal control sequence $\bm{u}^*(k) = \left\{u^*(k \mid k), \dots ,u^*(k+N-1 \mid k) \right\}$. $\mathbb{X}_N$ is assumed to be a closed subset of $\mathbb{X}$, and the origin is in its interior. Besides, it is assumed that for all $x \in \mathbb{X}_N$, there exists an implicit control law $u_N(x) \in \mathbb{U}$ such that $f(x,u_N(x))\in\mathbb{X}_N$ and $V_f(f(x,u_N(x))\le V_f(x) - l(x,u_N(x))$. Additionally, to prove the recursive feasibility of the proposed MPC, suppose that the optimization problem \eqref{equation:ocp_proof} is feasible at initial time. At time instant $k$, the optimal control $\bm{u}^*(k)$ can steer current state $x(k)$ to the optimal state trajectory $\bm{x}^*(k) = \{x^*(k \mid k), \dots ,x^*(k+N \mid k)\}$ where $x^*(k+N \mid k) \in \mathbb{X}_N$. At next time instant $k+1$, a control sequence is constructed by 
\begin{equation}\label{equation:u_in_k+1}
        \tilde{\mathsf{u}}(k+1)=\left\{\begin{array}{ll}
                u^*(k+i \mid k), & i =1, \dots, N-1, \\
                u_{N}\left(x^*(k+N \mid k)\right), & i=N .
                \end{array}\right.
\end{equation}

Using the constructed control sequence \eqref{equation:u_in_k+1} at time instant $k+1$, the state sequence $x(k+i \mid k+1), i=1,\dots,N $ can be obtained, where $x(k+N \mid k+1) \in \mathbb{X}_N $. Under the control command $u_{N}\left(x^*(k+N \mid k)\right)$ in \eqref{equation:u_in_k+1}, the state $x(k+N+1 \mid k+1) = f(x^*(k+N \mid k),u_N(x^*(k+N|k)) \in \mathbb{X}_N$ is feasible. Therefore, the constructed control sequence \eqref{equation:u_in_k+1} is a feasible solution of the proposed MPC at time instant $k+1$, which means that the recursive feasibility of the proposed MPC is ensured.

Then the stability of the closed-loop system with the proposed MPC is established, with a Lyapunov function candidate defined as $V_N^0(x(k)) = \min\limits_{\bm{u}} \sum_{i=0}^{N-1}l(x^*(k+i \mid k), u^*(k+i \mid k)) + V_f(x^*(k+N \mid k))$, at time instant $k$. Considering the constructed feasible control sequence \eqref{equation:u_in_k+1}, the Lyapunov function $V_N^0(x(k+1))$ at next time instant $k+1$ can be written as 

\begin{equation}
        \begin{aligned}
        &V_N^0(x(k+1))\\
        =& \min\limits_{\bm{u}} \sum_{i=0}^{N-1}l(x^*(k+1+i \mid k+1), u^*(k+1+i \mid k+1))\\
        & + V_f(x^*(k+1+N \mid k+1))\\
        \le& V_N^0(x(k)) - l(x^*(k \mid k),u^*(k \mid k)).
        \end{aligned}
\end{equation}

Therefore, $V_N^0(x(k+1)) < V_N^0(x(k))$ holds until the state $x(k)$ converges to zero, and the stability of the closed-loop system with the proposed MPC is established.

\subsubsection{INDI Angular Velocity Controller}
To track the angular velocity command $\bm{\omega}_{c}$ in the presence of unknown torque disturbance $\bm{\tau}_d$, an INDI angular velocity controller is employed in this section \cite{ref6}. Consider quadrotor dynamics (\ref{tau_d}), which can be rewritten as the following expression:

\begin{equation}\label{equation:innerDO}
        \bm{\tau}_d = \bm{J}\dot{\bm{\omega}}_{f}-\bm{\tau}_{c,f}+\bm{\omega}_{f}\times\bm{J}\bm{\omega}_{f},
\end{equation}
where $\dot{\bm{\omega}}_{f}$ is the feedback angular acceleration, $\bm{\tau}_{c,f}$ is the torque command from the last computation, and ${\bm{\omega}}_{f}$ represents the feedback angular velocity. Substituting (\ref{equation:innerDO}) into (\ref{tau_d}) yields:
\begin{equation}\label{equation:INDI}
        \begin{aligned}
                \dot{\bm{\omega}}&=\bm{J}^{-1}[\bm{\tau}_c-\bm{\omega}\times\bm{J}\bm{\omega}]+\bm{J}^{-1}\bm{\tau}_d\\
                &=\bm{J}^{-1}[\bm{\tau}_c-\bm{\omega}\times\bm{J}\bm{\omega}+\left(\bm{J}\dot{\bm{\omega}}_{f}-\bm{\tau}_{c,f}+\bm{\omega}_{f}\times\bm{J}\bm{\omega}_{f}\right)]\\
                &=\dot{\bm{\omega}}_{f}+\bm{J}^{-1}[\bm{\tau}_c-\bm{\tau}_{c,f}],
        \end{aligned} 
\end{equation}
where it is supposed that $\bm{\omega}\times\bm{J}\bm{\omega}=\bm{\omega}_{f}\times\bm{J}\bm{\omega}_{f}$, since this term is regarded as relatively slow-changing in comparison to the angular acceleration $\dot{\bm{\omega}}$ and the torque command $\bm{\tau}$ \cite{ref6,ref6_4}. Then, the control command $\bm{\tau}_c$ can be calculated by
\begin{equation}
        \label{equation:tau}
        \bm{\tau}_c = \bm{\tau}_{c,f}+\bm{J}(\dot{\bm{\omega}}_{c}-\dot{\bm{\omega}}_{f}),
\end{equation}
where $\dot{\bm{\omega}}_{c}$ is the virtual angular acceleration command which can be derived from the following proportional controller:
\begin{equation}
        \label{equation:omega_c}
        \dot{\bm{\omega}}_{c} = \bm{K}_{\omega}(\bm{\omega}_{c}-\bm{\omega}_f) + \dot{\bm{\omega}}_r,
\end{equation}
where $\bm{\omega}_{c}$ is the reference angular velocity that comes from the control command of the FxTDO-based MPC controller, and $\dot{\bm{\omega}}_r$ represents the reference angular acceleration.

\begin{Rem}
        In the real-world quadrotor platform, control commands sent to quadrotor need to be computed at a very high frequency, such as 1000 Hz \cite{ref6}. However, the excessive computational time required of MPC on the onboard computer poses a limitation on its applicability within the inner-loop. Therefore, it is necessary to incorporate an inner-loop controller with lower computational requirements. Considering the advantages of INDI method, including the low computational requirements and sufficient robustness \cite{ref6,10243043}, an INDI angular velocity controller is employed as the inner-loop controller.
\end{Rem}

\section{Simulation and Experiment}
In this section, simulations in Gazebo and real-world experiments are developed to evaluate the effectiveness of the proposed method. For brevity, the proposed method is denoted as FxTDO-MPC \& INDI. To highlight the effectiveness of the proposed method, a conventional PID controller, a conventional MPC controller, a robust tube-based MPC (RT-MPC) controller \cite{mayne2005robust} and a high-gain disturbance observer-based MPC (HGDO-MPC) controller \cite{khalil2014high} are employed for comparison, incorporating with inner-loop INDI angular velocity controller. The MPC, RT-MPC and HGDO-MPC methods adopt the same set of parameters as the FxTDO-MPC method, while the PID controller is carefully tuned for optimal performance. Notably, the RT-MPC framework proposed in \cite{mayne2005robust} is designed for linear systems. To adopt the RT-MPC method for quadrotor, several modifications are implemented as follows. Firstly, the nominal MPC controller within the RT-MPC framework is constructed based on a nonlinear model regardless of disturbances, generating optimal control sequence $\bm{u}^*_n(k) = \left\{\bm{u}^*_n(k\mid k),\dots,\bm{u}^*_n(k+N-1\mid k) \right\}$ and the nominal optimal state sequence $\bm{x}^*(k) = \{ \bm{x}^*(k \mid k), \dots ,$ $ \bm{x}^*(k+N \mid k)\}$. Furthermore, this nominal MPC controller utilizes the control sequence and the initial state as decision variables \cite{mayne2005robust, mammarella2020tube}. Secondly, the quadrotor model is linearized into the discrete-time form: $\bm{x}_{k+1} = \bm{A}\bm{x}_{k} + \bm{B}\bm{u}_{k}$, and the feedback control law is defined: $\bm{u}_{f,k} = \bm{K}(\bm{x}_k - \bm{x}^*(k \mid k))$, where the gain matrix $\bm{K}$ is obtained by solving the associated Riccati equation, ensuring that $\bm{A}_K := \bm{A} + \bm{B}\bm{K}$ is stable \cite{mayne2005robust, mammarella2020tube}. Finally, by combining the nominal MPC controller and the feedback controller, the control command of RT-MPC $\bm{u}_{k}$ is computed as: $\bm{u}_{k} = \bm{u}^*_n(k \mid k) + \bm{K}(\bm{x}_k - \bm{x}^*(k \mid k)).$ On the other hand, an HGDO-MPC method is constructed by replacing FxTDO in the proposed FxTDO-MPC method with a multivariable HGDO. The HGDO takes the form: $\dot{\hat{\bm{z}}}_1 = \hat{\bm{z}}_2 + \bm{T} + \frac{\alpha_1}{\varepsilon}\bm{e}_1, \dot{\hat{\bm{f}}}_d = \frac{\alpha_2}{\varepsilon^2}\bm{e}_1,$ where $\alpha_1$, $\alpha_2$, and $\varepsilon$ are positive constants \cite{khalil2014high}.

\begin{figure}
        \centering
        \includegraphics[width=2.5in]{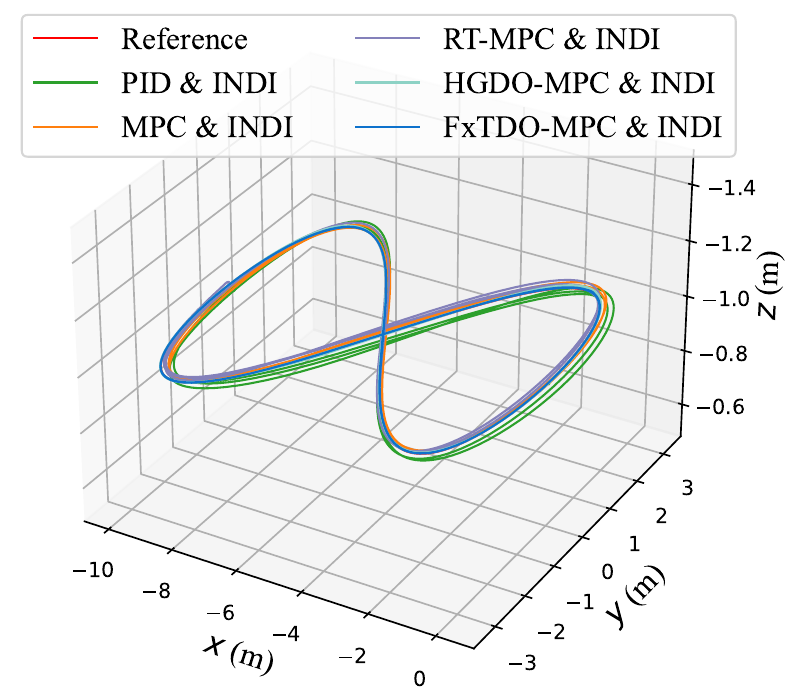}
        \caption{Quadrotor 3D trajectory tracking performance in Simulation.}
        \label{fig:Traj_Simulation}
\end{figure}

\begin{figure}
        \centering
        \includegraphics[width=3.3in]{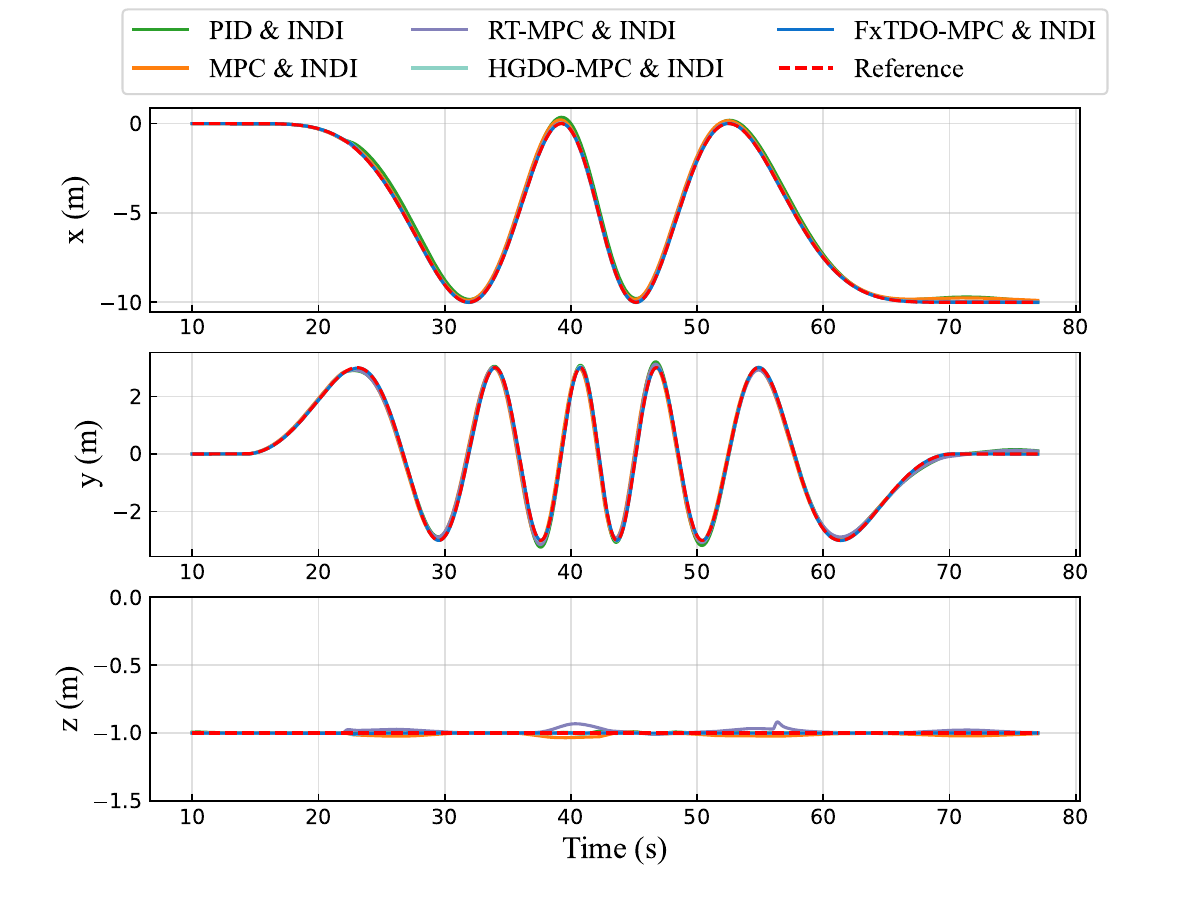}
        \caption{Position tracking performance in Simulation.}
        \label{fig:Position_Simulation}
        \vspace{-0.5cm}
\end{figure}

\subsection{Simulation Setups}
The proposed controller is executed in the ROS environment and sends control command $\left(T_c,\bm{\tau}_c\right)$ to quadrotor, which runs in Gazebo. The FxTDO-based MPC controller runs at 100 \text{Hz} while the INDI angular velocity controller operates at 1000 \text{Hz}. The model configurations for quadrotor in Gazebo are shown in Table \ref{table:model configurations}. The gains of the proposed controller and disturbance observer are summarized in Table~\ref{table:Controller parameters}.

In this scenario, a set of flight tests in simulation is developed to prove the effectiveness of the proposed controller and FxTDO. The desired trajectories $\bm{\varepsilon}_r$ considered here are the eight-shaped trajectories as shown in Fig.~\ref{fig:trajectory_bling}, which is defined as:
\begin{equation}
        \label{equation:traj}
        \bm{\varepsilon}_r=\left[\bm{p}_r,\psi_r\right]^\mathsf{T}=\begin{bmatrix}
         r_x\sin k_tt^2\cos k_tt^2\\
         r_y\cos k_tt^2-r_y\\
         r_z \\
        0 
        \end{bmatrix},
\end{equation}
with $r_x=3$ \text{m}, $r_y=5$ \text{m} ,$r_z=-1$ \text{m} and $k_t=0.01$. In the simulation, the complex disturbance effects $\bm{f}_{d}$ and $\bm{\tau}_d$ are defined:
\begin{equation}\label{equation:disturbance1}
        \bm{f}_{d} \!=\!\left[\begin{array}{c}
    1 + 0.5\sin \frac{2\pi\Delta t}{15} \\
    -0.5\cos \frac{2\pi\Delta t}{15} \\
    0
    \end{array}\right],
    \bm{\tau}_d\!=\!\left[\begin{array}{c}
            0.2\sin \frac{2\pi\Delta t}{15} \\
            0.2\cos \frac{2\pi\Delta t}{15} \\
            0
            \end{array}\right],
\end{equation}
where $\Delta t $ is the start time of injecting disturbances.

\begin{figure}
        \centering
        \includegraphics[width=3.3in]{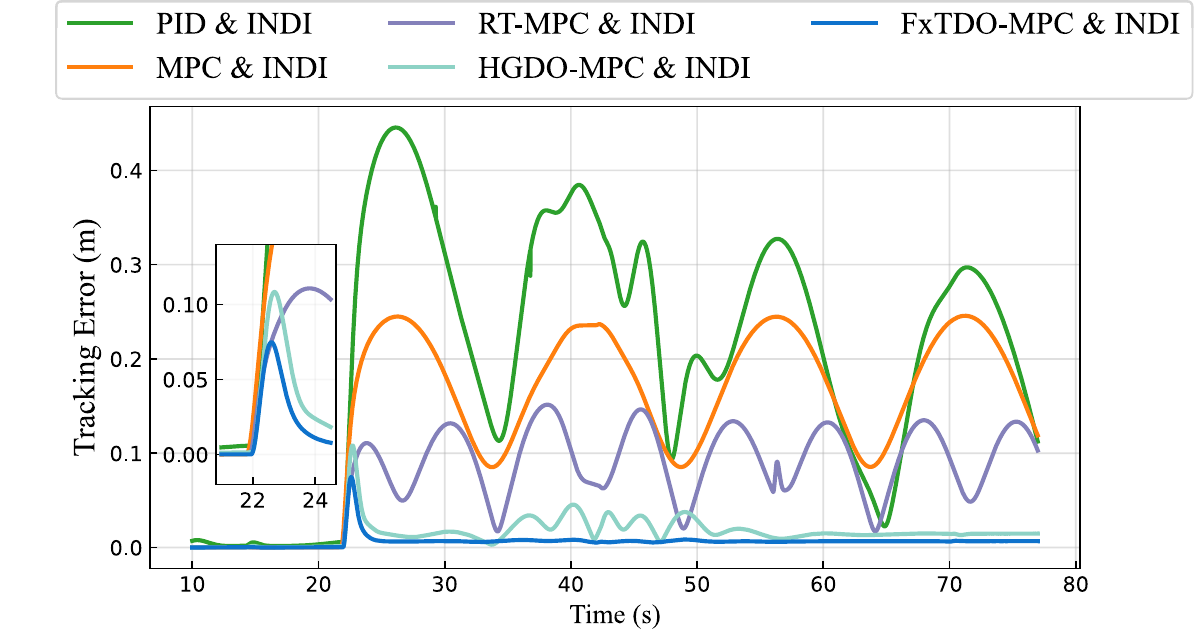}
        \caption{Position tracking error in Simulation.}
        \label{fig:Error_Simulation}
\end{figure}

\begin{figure}
        \centering
        \includegraphics[width=3.3in]{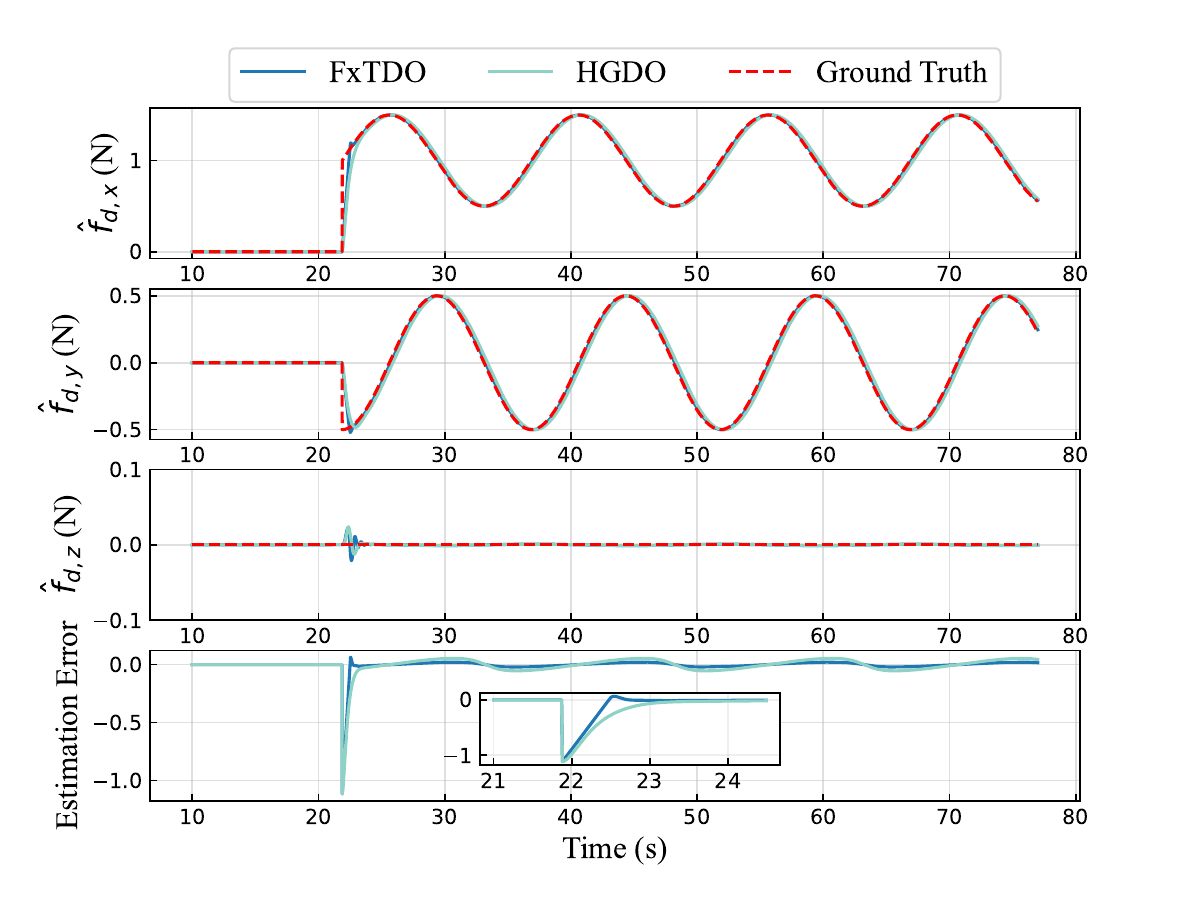}
        \caption{The disturbance estimation in Simulation.}
        \label{fig:do2_simulation}
        \vspace{-0.5cm}
\end{figure}

\subsection{Simulation Results}
\begin{figure}
        \centering
        \includegraphics[width=3.3in]{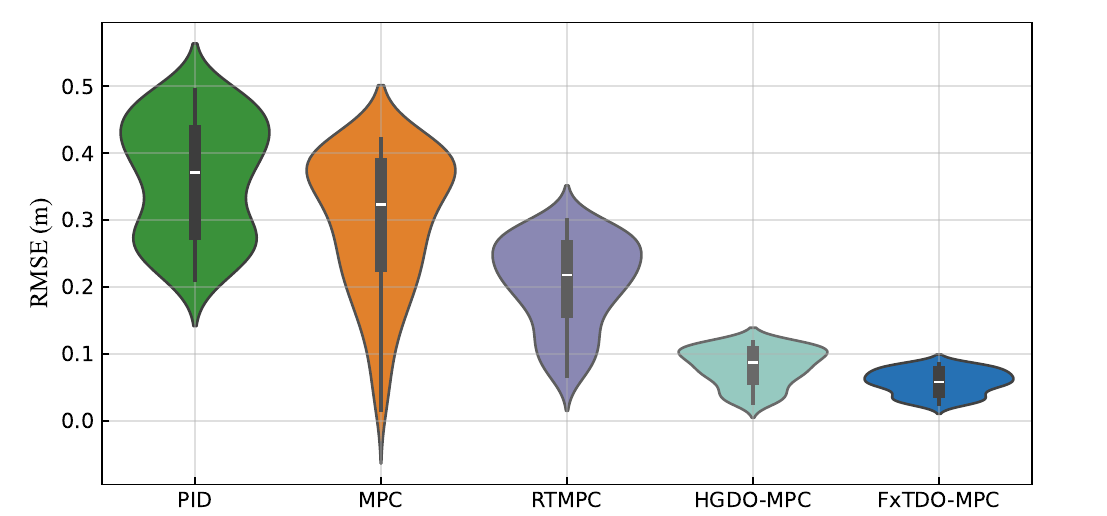}
        \caption{The distribution of position tracking RMSE of Monte Carlo simulations.}
        \label{fig:RMSE}
        \vspace{-0.5cm}
\end{figure}

The trajectory tracking performance in the influence of external disturbances \eqref{equation:disturbance1} is depicted in Figs.~\ref{fig:Traj_Simulation} and \ref{fig:Position_Simulation}, illustrating that the proposed method effectively enables the quadrotor to track the desired trajectory in the presence of disturbances. The position tracking error is presented in Fig.~\ref{fig:Error_Simulation}, with corresponding Root Mean Squared Errors (RMSE) of 0.245 m, 0.169 m, 0.092 m (RT-MPC \& INDI), 0.020 m (HGDO-MPC \& INDI), and 0.009 m (FxTDO-MPC \& INDI), respectively. It can be seen from Fig.~\ref{fig:Error_Simulation} that the proposed FxTDO-MPC \& INDI method achieves the lowest tracking error compared to the other methods. The disturbance estimation using two disturbance observers is depicted in Fig.~\ref{fig:do2_simulation}, illustrating that the proposed FxTDO converges within about 0.94 seconds. It is obvious that the proposed FxTDO achieves a faster convergence rate compared to the HGDO. The faster convergence of the FxTDO allows for quicker adjustments in the control inputs, leading to improved tracking performance, as illustrated in Fig.~\ref{fig:Error_Simulation}. Therefore, when subjected to disturbances, the method utilizing the proposed FxTDO exhibits a more rapid decrease in the position tracking error compared to the method employing the HGDO. Consequently, employing a disturbance observer with faster convergence properties in observer-based MPC scheme results in smaller transient error. In addition, to further validate the robust performance of the proposed method, Monte Carlo simulations are conducted by multiplying the disturbance $\bm{f}_d$ in \eqref{equation:disturbance1} by a random $k \in [0,1]$. Notably, the simulations are executed 500 times for all methods. The distribution of position tracking RMSE is summarized in Fig.~\ref{fig:RMSE}. It is obvious that the proposed method demonstrates the lowest error level with a concentrated error distribution, indicating the ability to ensure the robust performance of system in the presence of varying degrees of disturbances.


\begin{table}[tbp]
        \caption{Quadrotor Configurations in Simulation}
        \centering
        \label{table:model configurations}
        \begin{tabular}{c c | c c}
        \toprule
        Parameters & Values & Parameters & Values\\ 
        \midrule
        $m$ &  1.0 \text{kg} & $l$ &  0.17 \text{m}\\
        $J_x$ &  $2.64\times10^{-3}$ \text{kg}$\cdot{\text{m}}^2$ & $d_x$ &  $9.4\times10^{-9}$ {\text{m}}\\
        $J_y$ &  $2.64\times10^{-3}$ \text{kg}$\cdot{\text{m}}^2$ & $d_y$ &  $9.4\times10^{-9}$ {\text{m}}\\
        $J_z$ &  $4.96\times10^{-3}$ \text{kg}$\cdot{\text{m}}^2$ & $c_T$ &  $2.5\times10^{-9}$ {\text{m}}\\
        \bottomrule
        \end{tabular}
\end{table}

\begin{table}[tbp]
        \caption{The Parameter Values of the Adopted Method.}
        \centering
        \label{table:Controller parameters}
        \begin{tabular}{c c c}
        \toprule
        Methods & Parameters & Values \\ 
        \midrule
        \multirow{6}*{MPC}& $\bm{Q}_p$ /$\bm{P}_p$ &  $\text{diag}\left(1500,1500,1500\right)$\\
        & $\bm{Q}_v$ / $\bm{P}_v$&  $\text{diag}\left(400,400,400\right)$\\
        & $\bm{Q}_q$ / $\bm{P}_q$&  $\text{diag}\left(500,500,500,500\right)$\\
        & $\bm{R}$ &  $\text{diag}\left(1,10,10,10\right)$\\
        & $N$ &  10 \\
        & $dt$ &  100 {\text{ms}} \\
        \cmidrule(lr){2-3}
        INDI & $\bm{K}_\omega$ &  $\text{diag}\left(400,400,300\right)$ \\
        \cmidrule(lr){2-3}
        \multirow{5}*{FxTDO}& $k_i$ & 2.0 \\
        & $k'_i$ & 0.6 \\
        & $k''_i$ & 3.0 \\
        & $d_{\infty}$ & $1/3$ \\
        & $L_1, L_2$ & $1.0$ \\
        \cmidrule(lr){2-3}
        \multirow{3}*{HGDO}& $\alpha_1$ & 3.0 \\
        & $\alpha_2$ & 2.0 \\
        & $\varepsilon$ & 0.2 \\

        \bottomrule
        \end{tabular}
\end{table}

\subsection{Real-world Experiment Setups}
\begin{figure}
        \centering
        \includegraphics[width=3.2in]{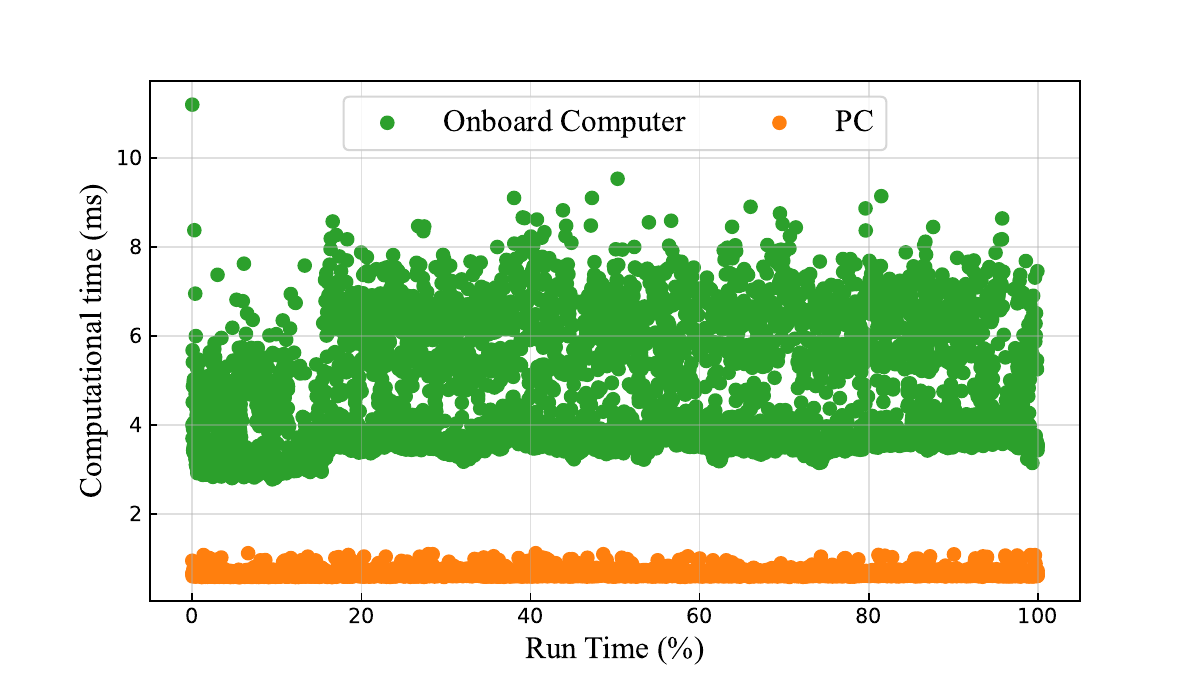}
        \caption{The computational time of the the proposed FxTDO-MPC method.}
        \label{fig:computational_time}
        \vspace{-0.5cm}
\end{figure}

To further verify the performance of the proposed method, a real quadrotor platform, as depicted in Fig.~\ref{fig:real quadrotor}, has been employed. The platform has a weight of 1.05 kg, with a thrust-to-weight ratio of 4:1. The inertia matrix $\bm{J}=\text{diag}\left(0.0235,0.0219,0.0285\right) \text{kg}{\text{m}}^2$ is roughly estimated through the bifilar pendulum experiment. The FxTDO-MPC algorithm runs on a Jetson Xavier NX onboard computer, computing the collective thrust and angular velocity. The computational time for each time step of the MPC on onboard computer and PC is depicted in Fig.~\ref{fig:computational_time}, where the PC is equipped with an Intel Core i9-11900, an NVIDIA GeForce RTX 3070 and 32G RAM. The INDI angular velocity controller operates on a CUAV v5+ autopilot running the PX4 firmware. The gains and parameters used in real-world experiments are the same as those in Table \ref{table:Controller parameters}. All real-world flight experiments are performed under the Optitrack motion capture system. In addition, the PX4 firmware provides the extended Kalman filter for state estimation which fuses Inertial Measurement Unit (IMU) data with pose information from Optitrack. Two experiment scenarios with disturbances are tested: \textit{1) Hovering performance with persistent wind disturbance (Scenario 1)}; \textit{2) Trajectory tracking with complex disturbances (Scenario 2)}.

\subsection{Real-world Experiment Results}
\subsubsection{Hover performance with persistent wind disturbance (Scenario 1)}
\begin{figure}
        \centering
        \includegraphics[width=3.2in]{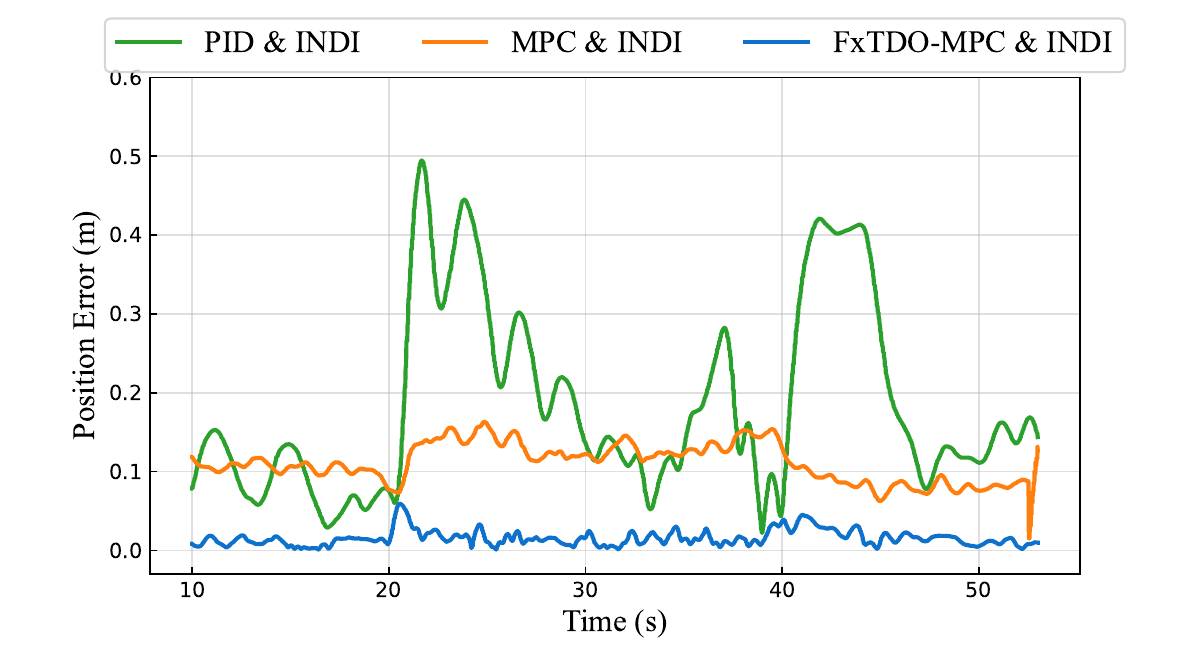}
        \caption{Position error in \textit{Scenario 1}.}
        \label{fig:Error_Hover}
        \vspace{-0.5cm}
\end{figure}

\begin{figure}
        \centering
        \includegraphics[width=3.2in]{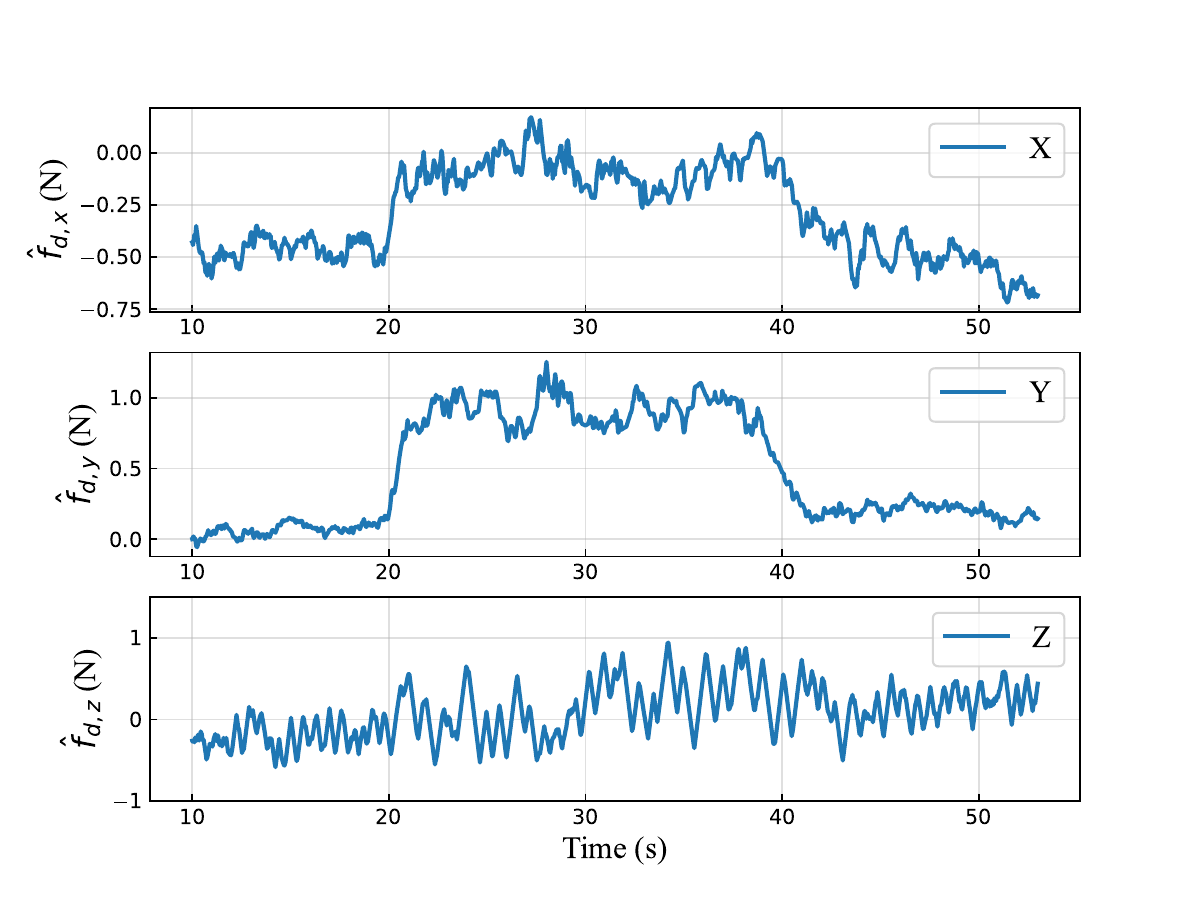}
        \caption{The disturbance estimation in \textit{Scenario 1}.}
        \label{fig:DO_Hover}
        \vspace{-0.5cm}
\end{figure}

In this scenario, a quadrotor hovers in a fixed position. Persistent wind gusts are generated by an electric fan. The wind speed is measured using an anemometer in the position where the quadrotor hovers, which is about 5.0 m/s, as shown in Fig.~\ref{fig:disturbance_setup}. The quadrotor platform exhibits model uncertainty primarily due to the inaccurate inertia matrix. The position error and RMSE are summarized in Fig.~\ref{fig:Error_Hover} and Table \ref{table:RMSE comparison}. From these results, it can be seen that the proposed method achieves a remarkably high level of control accuracy, leading to a large reduction in RMSE compared to the MPC without FxTDO and classical PID controller. The results of the FxTDO are provided in Fig.~\ref{fig:DO_Hover}. The electric fan is activated at around 20 seconds, and it can be observed that the proposed FxTDO is able to quickly estimate the disturbance mainly caused by the electric fan.

\begin{table}[tbp]
        \caption{The position tracking RMSE comparison in real-world experimets (Scenario 1 and Scenario 2).}
        \label{table:RMSE comparison}
        \centering
        \begin{tabular}{c c c}
        \toprule
            Methods & \multicolumn{2}{c}{RMSE (m)}\\
        \midrule
           & \textit{Scenario 1} & \textit{Scenario 2}\\
         \cmidrule(lr){2-3}
          PID \& INDI  &  0.219  & 0.302\\
          MPC \& INDI &  0.113  & 0.178\\
          FxTDO-MPC \& INDI  & \textbf{0.018} & \textbf{0.051} \\
        \bottomrule
\end{tabular}
\end{table}


\begin{figure}
        \centering
        \includegraphics[width=2.5in]{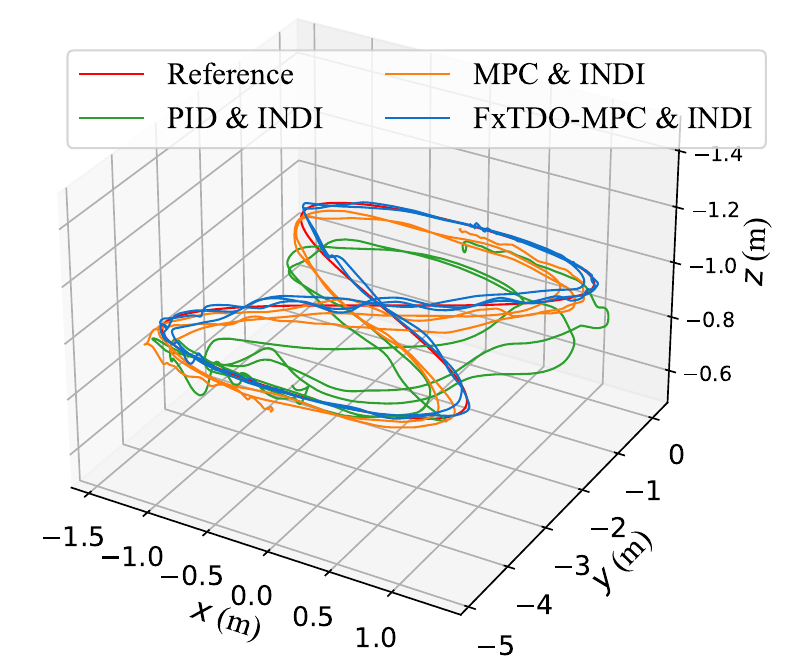}
        \caption{Quadrotor 3D trajectory tracking performance in \textit{Scenario 2}.}
        \label{fig:Traj_Real}
\end{figure}

\begin{figure}[htb]
        \centering
        \includegraphics[width=3.3in]{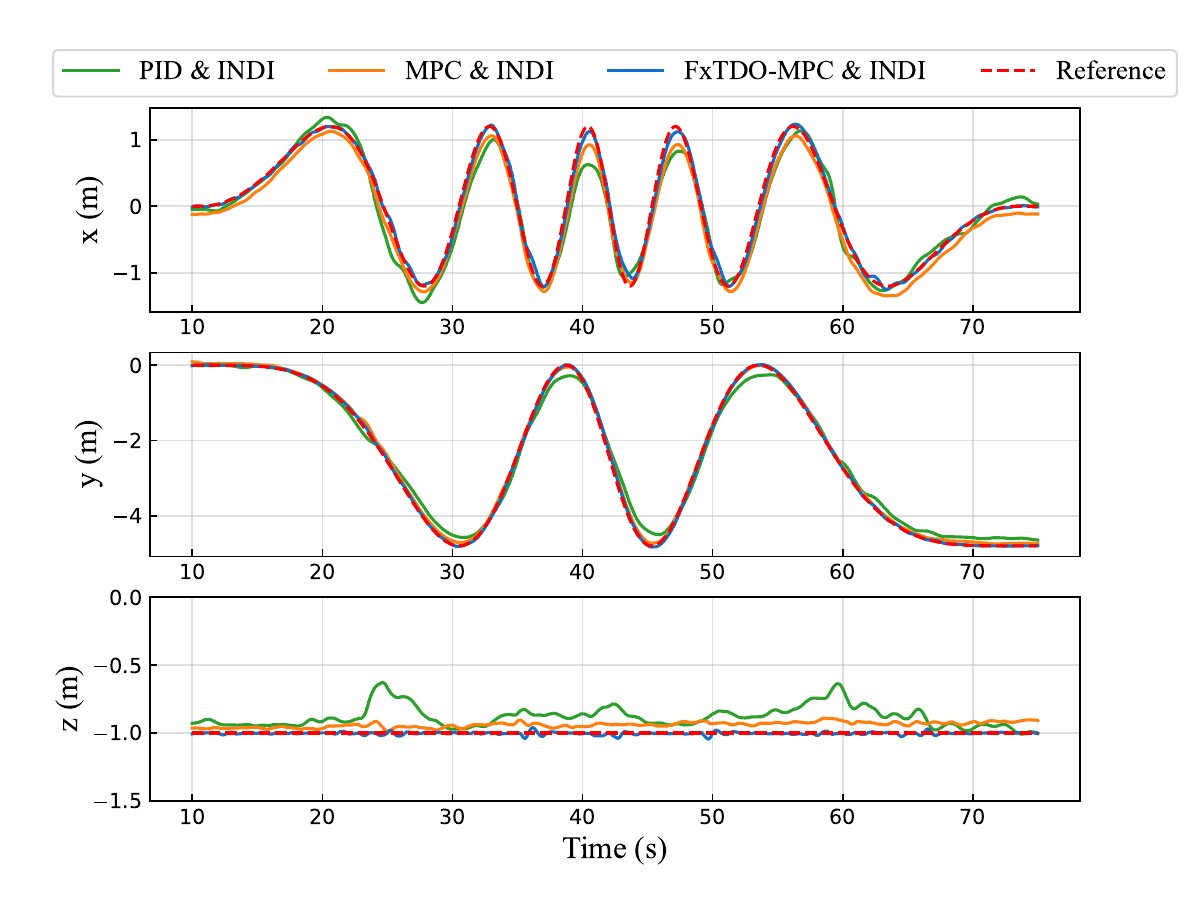}
        \caption{Position tracking performance in \textit{Scenario 2}.}
        \label{fig:Position_Real}
        \vspace{-0.5cm}
\end{figure}

\subsubsection{Trajectory tracking with complex disturbances (Scenario 2)}
\begin{figure}[htb]
        \centering
        \includegraphics[width=3.2in]{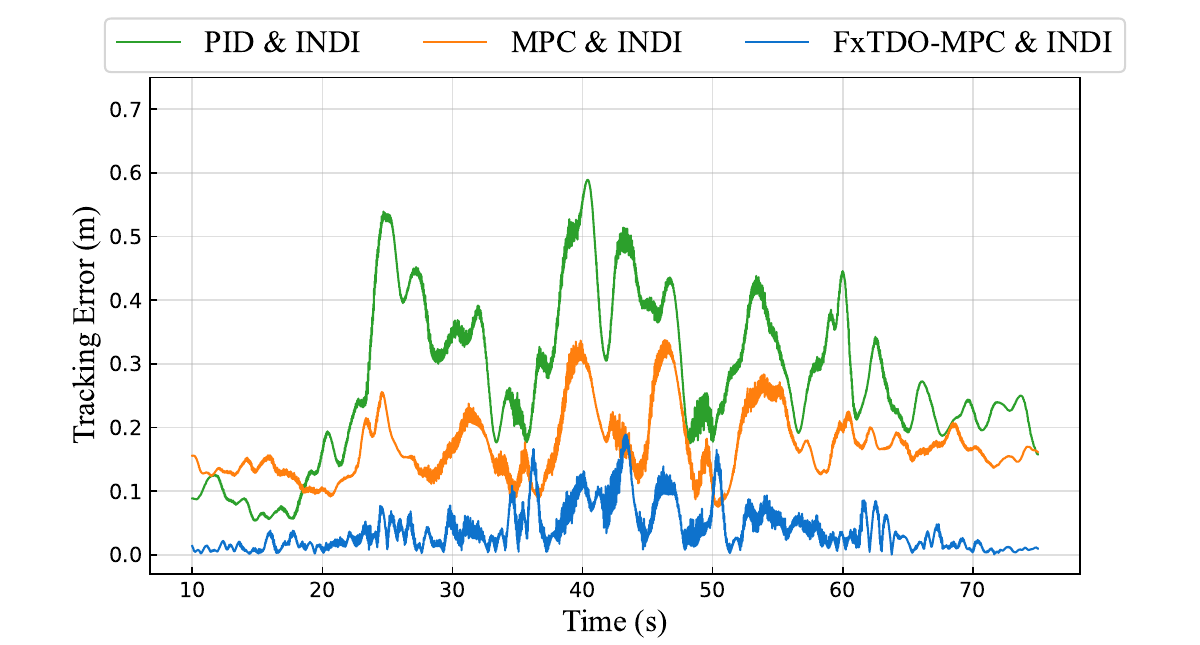}
        \caption{Position tracking error in \textit{Scenario 2}.}
        \label{fig:Error_Traj}
        \vspace{-0.5cm}
\end{figure}

\begin{figure}[htb]
        \centering
        \includegraphics[width=3.2in]{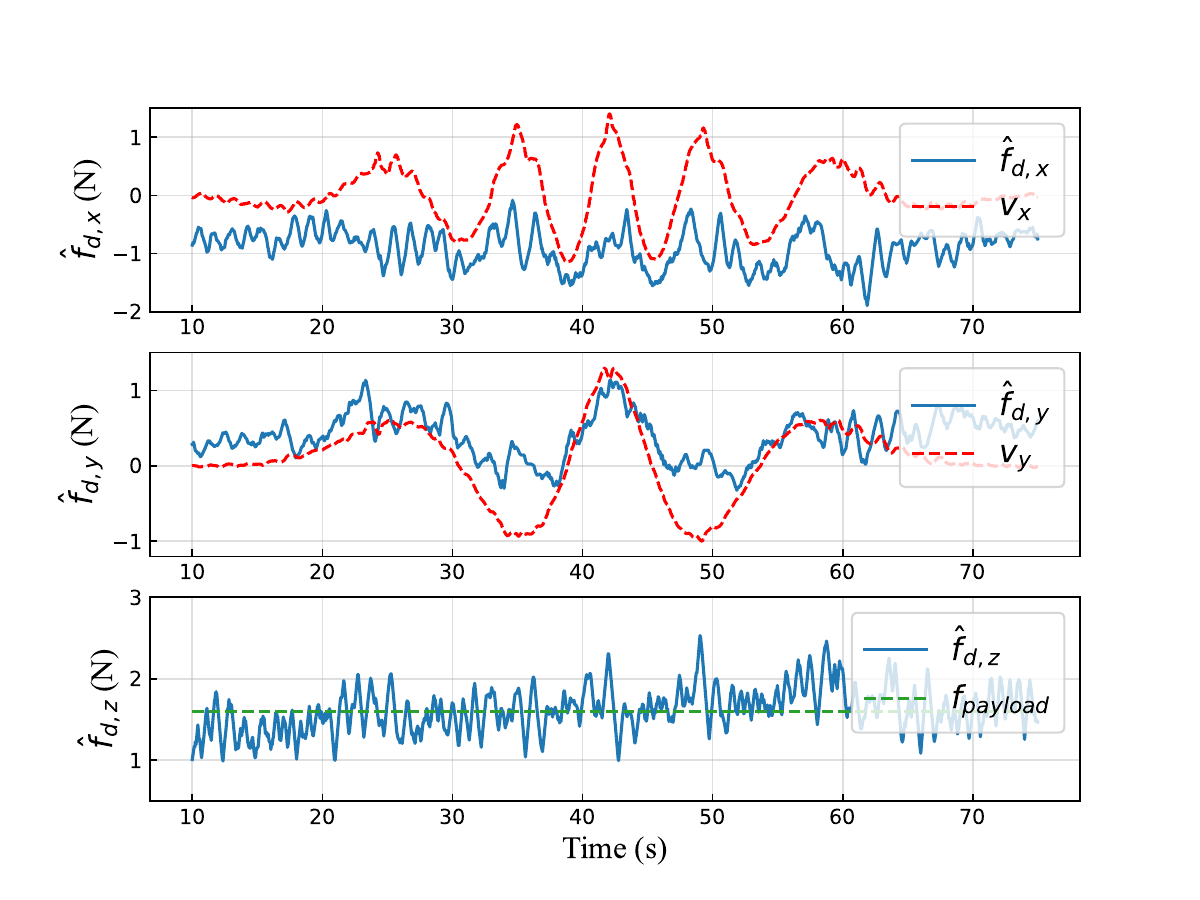}
        \caption{The disturbance estimation in \textit{Scenario 2}.}
        \label{fig:DO_Traj}
        \vspace{-0.5cm}
\end{figure}

In order to assess the efficiency of the proposed algorithm in the presence of complex disturbances, a payload with a mass of 163 g is affixed to the quadrotor, while wind disturbance with speed ranging from 3.0 m/s to 4.8 m/s is introduced in the center of the experimental site, as shown in Fig.~\ref{fig:disturbance_setup}. Due to the limited space in the indoor environment, the desired trajectory defined in (\ref{equation:traj}) with $r_x=1.2$ \text{m}, $r_y=2.4$ \text{m}, $r_z=1.0$ \text{m} and $k_t=0.01$ is used. Figs.~\ref{fig:Traj_Real} and \ref{fig:Position_Real} illustrate the trajectory tracking performance under three different methods. Furthermore, the tracking error can be found in Fig.~\ref{fig:Error_Traj} and Table~\ref{table:RMSE comparison}, where the proposed method achieves 72\% and 83\% decrease in RMSE compared to MPC \& INDI and PID \& INDI respectively. From the position error in Fig.~\ref{fig:Error_Traj} and RMSE in Table~\ref{table:RMSE comparison}, it can be concluded that the proposed approach achieves accurate trajectory tracking while effectively rejecting complex disturbances. Fig.~\ref{fig:DO_Traj} presents the estimated value of FxTDO in \textit{Scenario 2}, which includes the complex influence of swaying load, electric fan, model uncertainty, and aerodynamic drag force during flight. For $\hat{f}_{d,x}$ and $\hat{f}_{d,y}$ in \textit{Scenario 2}, the primary source of disturbance is the electric fan and aerodynamic drag force which is related to the velocity of quadrotor \cite{ref6_4}. The blue line in Fig.~\ref{fig:DO_Traj} represents the estimated value of FxTDO, while the red dashed line represents the velocity of quadrotor. Although these two variables do not exhibit a linear correlation, they demonstrate a consistent similarity in the overall pattern of change. It should be noted that the green dashed line represents the force caused by the payload, and the estimated value $\hat{f}_{d,z}$ demonstrates fluctuation around the green dashed line, which is attributed to the swaying motion of the payload.

\section{Conclusion}
This paper proposes an FxTDO-MPC algorithm for robust trajectory tracking of quadrotor in the presence of disturbances. Firstly, the FxTDO is introduced to estimate the lumped disturbances. The convergence of estimation error within a fixed convergence time is guaranteed by the bi-limit homogeneity and Lyapunov techniques. Then, the observer-based model predictive controller is formulated by integrating the estimation of FxTDO into the prediction model. The proposed method achieves accurate trajectory tracking and robust disturbance rejection of quadrotor, while simulations and real-world experiments are developed to evaluate the effectiveness of the proposed method. Future work will focus on extending the proposed method to various vehicle models, such as fixed-wing UAVs, and to diverse operational environments, such as scenarios involving actuator failures. 

\bibliographystyle{IEEEtran}
\bibliography{arxiv}

\end{document}